\documentclass[]{JFM-FLM_Au}

\usepackage{physics}
\usepackage{multirow}
\usepackage{tikz}
\usepackage{xcolor}
\usepackage{siunitx}
\definecolor{mp01}{RGB}{110,75,190}   
\definecolor{mp02}{RGB}{45,165,175}   
\definecolor{mp04}{RGB}{140,205,90}   
\definecolor{mp06}{RGB}{220,205,70}   

\def\d{\mathrm{d}}
\def\iunit{\mathrm{i}}
\def\average#1{\left\langle{#1}\right\rangle}
\def\averagespace#1{\left\langle{#1}\right\rangle_{\mathcal{V}}}
\def\averagetime#1{\left\langle{#1}\right\rangle_t}
\def\averagespacetime#1{\left\langle{#1}\right\rangle_{\mathcal{V},t}}
\def\averageparticle#1{\left\langle{#1}\right\rangle_p}
\def\averageparticletime#1{\left\langle{#1}\right\rangle_{p,t}}
\def\Stokes{\text{\textit{St}}}
\def\urms{u_{\text{rms}}}
\def\particle{\text{part}}
\def\collision{\text{col}}
\def\shell{\text{shell}}
\def\averageshell#1{\left\langle{#1}\right\rangle_{\shell}}
\def\vvoro{\mathcal{V}_{\text{voro}}}
\def\sigmavoro{\sigma_{\text{voro}}}
\def\skewvoro{\mathcal{S}_{\text{voro}}}
\def\kurtvoro{\mathcal{K}_{\text{voro}}}
\def\Ncluster{N_{\text{cluster}}}
\def\Vcluster{\mathcal{V}_{\text{cluster}}}
\def\ABC{\text{ABC}}
\def\levicivita{\epsilon_{ijk}}

\newcommand{\bluealert}[1]{\textcolor{blue!85!black}{\textbf{#1}}}

\DeclareRobustCommand{\trianglemark}{%
  \leavevmode
  \tikz[baseline=-0.7ex]\draw[fill=none]
  (0,0.1)--(0.1,-0.1)--(-0.1,-0.1)--cycle;
}
\DeclareRobustCommand{\diamondmark}{%
  \leavevmode
  \tikz[baseline=-0.7ex]\draw[fill=none] (0,0.1) -- (0.1,0) -- (0,-0.1) -- (-0.1,0) -- cycle;
}
\DeclareRobustCommand{\circlemark}{%
  \leavevmode
  \tikz[baseline=-0.7ex]\draw[fill=none]
  (0,0) circle (0.1);
}

\lefttitle{L. Foss\`{a} and M. E. Rosti}
\righttitle{Journal of Fluid Mechanics}

\title{A suspension of heavy Kolmogorov-size spheres suppresses the inertial cascade in homogeneous and isotropic turbulence}

\author{Ludovico Foss\`{a} \and Marco Edoardo Rosti}

\affiliation{Complex Fluids and Flows Unit, Okinawa Institute of Science and Technology Graduate University, 1919-1 Tancha, Onna-son, Kunigami-gun, Okinawa-ken 904-0495 Japan}

\corresau{Marco Edoardo Rosti \email{marco-rosti@oist.jp}}

\begin{document}
\maketitle

\begin{abstract}
\bluealert{Accepted for publication in Journal of Fluid Mechanics}

The effect of Kolmogorov-size spherical particles on homogeneous and isotropic turbulence is investigated using particle-resolved direct numerical simulations at an unladen Taylor-scale Reynolds number of $150$. Four monodisperse suspensions of particles with identical diameter and volume fraction $10^{-3}$ are considered, while the particle-to-fluid density ratio varies between $100$ and $1500$ and the mass fraction between $0.1$ and $0.6$. As particle inertia increases, the energy spectrum departs from the canonical Kolmogorov $\kappa^{-5/3}$ scaling and approaches a peculiar regime with  $\kappa^{-1}$. In this limit, the nonlinear energy transfer is strongly suppressed and the kinetic energy balance is dominated by the fluid-solid interaction and the viscous dissipation. Consistently, the second-order structure function shows logarithmic scaling at separations larger than the particle diameter, indicating velocity decorrelation. Increasing particle inertia promotes axial strain and vortex compression in the vicinity of the particles and enhances the particle-fluid relative velocity. Particle clustering is maximum when the Stokes number based on the Kolmogorov time scale is $O(1)$ and weakens as the density ratio and the Stokes number increase, with the volume and the population of the clusters decreasing when inertia is enhanced. When clustering occurs, particles preferentially sample regions of high strain and low vorticity.
\end{abstract}



\section{Introduction}
\label{sec:intro}

Particle-laden flows are ubiquitous in nature and engineering \citep{Brandt_Coletti_2022,Marchioli_Bourgoin_Coletti_Fox_Magnaudet_Reeks_Simonin_Sommerfeld_Toschi_Wang_Balachandar_2025,Marchioli_Rosti_Verhille_2025}. Examples include cloud droplets and volcanic ashes in planetary atmospheres \citep{Mellado_2017,Rose_Durant_2009}, the formation of protoplanetary disks \citep{Birnstiel_2024}, the dispersal of phytoplankton and microplastics in the ocean \citep{Borgnino_Arrieta_Boffetta_De-Lillo_Tuval_2019,Sugathapala_Capuano_Brandt_Iudicone_Sardina_2025}, the transport of blood cells \citep{Freund_2014} and industrial processes such as flue gas scrubbing \citep{Lee_Raj-Mohan_Byeon_Lim_Hong_2013} and fuel atomization in spray combustion \citep{Jenny_Roekaerts_Beishuizen_2012}. These phenomena see a turbulent carrier flow transporting and interacting with a solid (or liquid) suspension that can either energise or weaken its motion. The nature of this modulation depends on factors such as the relative size of the particle and the turbulent eddies \citep{Gore_Crowe_1989}, as well as the inertia and degree of dilution of the suspension \citep{Balachandar_Eaton_2010}. A suspension is considered dilute or dense in the limit of low or high volume fraction $\Phi_p = \mathcal{V}_p/(\mathcal{V}_f + \mathcal{V}_p)$, where $\mathcal{V}_f$ and $\mathcal{V}_p$ denote the volumes occupied by the fluid and dispersed phases, respectively. The threshold also depends on the density ratio $\Psi_p = \rho_p/\rho_f$, where $\rho_p$ and $\rho_f$ are the particle and fluid densities, respectively, and on the particle size size $D$ \citep{Elghobashi_1994,Balachandar_Eaton_2010}. In general, when $D$ is much smaller than the Kolmogorov length scale $\eta$, the motion of the particles is primarily governed by that of the local fluid, the strength of their response being modulated by inertia. Particles of negligible inertia behave as pure tracers and follow the motion of the carrier flow, while particles of large inertia are only weakly affected by the surface stresses induced by the carrier flow and do not follow its streamlines. Particles of moderate inertia are expected to be only partially influenced by the carrier flow, their trajectory departing from its streamlines in the presence of intense velocity gradients. This centrifugal mechanism was proposed by \cite{Maxey_1987}, who showed that small, light spheres segregate in regions of low vorticity and high strain, and that this behaviour intensifies for increasing particle mass. His predictions were later confirmed numerically by \cite{Squires_Eaton_1990}. This preferential sampling of the flow has been indicated as a key process in the formation and evolution of clusters in homogeneous and isotropic turbulence (see the reviews of \citealp{Balachandar_Eaton_2010} and \citealp{Brandt_Coletti_2022}). The sweep-stick mechanism of \cite{Goto_Vassilicos_2008} has also been proposed to account for preferential sampling in flows laden with smaller-than-Kolmogorov and finite-size particles \citep{Yoshimoto_Goto_2007,Uhlmann_Chouippe_2017}. 

When forming dilute suspensions, low-inertia, smaller-than-Kolmogorov particles behave essentially as passive tracers of a one-way coupled mechanical system \citep{Volk_Calzavarini_Verhille_Lohse_Mordant_Pinton_Toschi_2008,Sumbekova_Cartellier_Aliseda_Bourgoin_2017}. These flows have been extensively investigated with the aid of point-particle direct numerical simulations (DNSs) where the motion of each particle is governed by a Lagrangian tracker \citep{Maxey_Riley_1983,Matsuda_Yoshimatsu_Schneider_2024}. Point-particle DNSs rely on the assumption of uniform flow properties across larger-than-particle distances $D/\eta\ll1$, meaning that the sub-Kolmogorov scales need not be resolved \citep{Brandt_Coletti_2022}. Although enormous efforts have been made in the development of two-way coupling models \citep{Balachandar_Liu_Lakhote_2019}, point-particle DNSs are yet to provide an accurate and complete description of the particle-fluid interaction when $D/\eta=O(1)$, a condition typical of several practical scenarios in which the dispersed phase modulates the turbulence \citep{Luo_Wang_Li_Tan_Fan_2017,Tanaka_Eaton_2010,Oka_Goto_2022}. In recent years, particle-resolved direct numerical simulations (PR-DNSs) based on the immersed boundary method (IBM) have cast new light on the interphase coupling in suspensions of larger-than-Kolmogorov ($D/\eta\gg1$) spherical particles \citep{Kajishima_Takiguchi_Hamasaki_Miyake_2001,Uhlmann_2005,Breugem_2012,Kempe_Frohlich_2012,Hori_Takagi_Rosti_2022}. To name a few, we recall the works by \cite{Lucci_Ferrante_Elghobashi_2010}, \cite{Uhlmann_Chouippe_2017}, \cite{Chouippe_Uhlmann_2019}, \cite{Oka_Goto_2022}, \cite{Olivieri_Cannon_Rosti_2022}, \cite{Chiarini_Rosti_2024} and \cite{Cannon_Olivieri_Rosti_2024} for homogeneous isotropic turbulence, and those by \cite{Uhlmann_2008}, \cite{Ardekani_Brandt_2019}, \cite{Rosti_Brandt_2020}, 
\cite{Costa_Brandt_Picano_2021} and \cite{Gao_Samtaney_Richter_2023} for turbulent channel flows. These works highlight the importance of the ratio $D/\eta$ on turbulence modulation: inertial-scale particles provide a spectral shortcut by extracting energy from the large scales and reinjecting it into the small ones \citep{Lucci_Ferrante_Elghobashi_2010,Oka_Goto_2022,Chiarini_Rosti_2024}, whereas Kolmogorov-size particles absorb the carrier flow's kinetic energy \citep{Tanaka_Eaton_2010} and may impose a decay rate of the energy spectrum proportional to the inverse fourth power of their spatial wavenumber \citep{Wang_Ayala_Gao_Andersen_Mathews_2014,Chiarini_Tandurella_Rosti_2025}. The exact conditions under which turbulence enhancement or attenuation occur remain unclear \citep{Marchioli_Bourgoin_Coletti_Fox_Magnaudet_Reeks_Simonin_Sommerfeld_Toschi_Wang_Balachandar_2025} and likely result from a nontrivial interplay between several parameters \citep{Tanaka_Eaton_2008}. In the absence of gravity, the PR-DNSs of \cite{Lucci_Ferrante_Elghobashi_2010} reported turbulence attenuation in triperiodic homogeneous isotropic turbulence laden with suspensions of Taylor-scale spheres with $0.01 \leq \Phi_p \leq 0.1$ and $2.56 \leq \Psi_p \leq 10$, whereas \cite{Chiarini_Rosti_2024} observed a non-monotonic dependence of the kinetic energy on $M_p$ and $D/\eta$ for $M_p \geq 0.45$ and $D/\eta \leq 64$. In the latter regime, the enhancement of turbulent kinetic energy was associated with a large-scale anisotropic modulation \citep{Chiarini_Cannon_Rosti_2024}. Turbulence enhancement has also been observed experimentally in settling suspensions of heavy sub-Kolmogorov particles with higher density ratios $\Psi_p = O(10^3)$ \citep{Yang_Shy_2005} and has been attributed to the release of gravitational potential energy \citep{Hassaini_Coletti_2022}. 

Small, heavy particles can attenuate the turbulent fluctuations just like large, light ones \citep{Oka_Goto_2022}. Along with the volume fraction $\Phi_p$ and the size ratio $D/\eta$, the particle-to-fluid density ratio $\Psi_p$ plays a key role on the interphase coupling \citep{Brandt_Coletti_2022}. \cite{Yu_Lin_Shao_Wang_2017} performed PR-DNSs with a direct forcing-fictitious domain method to study the effect of increasing $\Psi_p$ in the range $1-100$ on a turbulent channel flow at a friction Reynolds number $\Rey_\tau=180$ and volume fraction $\Phi_p=8.4\times10^{-3}$. They considered two sets of spherical particles of diameter $0.1$ and $0.2$ times the channel half-width, which corresponds to $D/\eta\cong5$ and $D/\eta\cong10$, respectively. They reported a non-monotonical trend of the total drag, which first increased along with the viscous dissipation for $\Psi_p\leq 10$, but diminished as the formation of near-wall streamwise vortices was inhibited \citep{Lucci_Ferrante_Elghobashi_2010}. \cite{Tavanashad_Passalacqua_Fox_Subramaniam_2019} investigated non-dilute populations of larger-than-Kolmogorov spherical particles by performing PR-DNSs on a triperiodic box. They considered several orders of magnitude of $\Psi_p$ ($10^{-3}-10^{3}$) and showed that heavier suspensions extract more kinetic energy from the carrier flow. Although they did not report the Kolmogorov length scale nor the Reynolds number, they used two sets of particles with diameter $20$ and $30$ times larger than the grid cell. More recently, \cite{Shen_Peng_Lu_Wang_2024} performed PR-DNSs using a lattice-Boltzmann method to study the effect of variable particle density on finite-size spherical particles on a turbulent channel flow, with particles in the range $6\lesssim D/\eta\lesssim 23$. Two-way coupled point-particle DNSs have also been employed to study the effect of particle inertia in suspensions of smaller-than-Kolmogorov spherical particles. \cite{Mortimer_Fairweather_2020} studied the effect of increasing $\Psi_p$ on near-wall coherent structures in a channel flow laden with $300\,000$ spheres at a friction Reynolds number $\Rey_\tau=180$ and volume fraction $\Phi_p=10^{-4}$. Their results showed a monotonical attenuation of the turbulent fluctuations in the near-wall region with increasing particle density. \cite{Gualtieri_Battista_Salvadore_Casciola_2023} conducted an extensive simulation campaign of non-dilute suspensions ($\Phi_p=0.4$) of smaller-than-Kolmogorov spheres in channel flows and concluded that dense particles deplete a considerable amount of turbulent kinetic energy. Consistently with the results of \cite{Yu_Lin_Shao_Wang_2017}, they reported an increase in friction drag in the presence of light particles, which acted as a source of turbulent kinetic energy near the wall.

Although predominantly studied numerically, particle-laden turbulent flows have also been the subject of several experimental studies. Most laboratory measurements have focused on suspensions of particles smaller than the Kolmogorov scale \citep[$D/\eta \ll 1$, e.g.][]{Yang_Shy_2005,Saw_Shaw_Ayyalasomayajula_Chuang_Gylfason_2008,Salazar_De-Jong_Cao_Woodward_Meng_Collins_2008,Obligado_Teitelbaum_Cartellier_Mininni_Bourgoin_2014,Good_Ireland_Bewley_Bodenschatz_Collins_Warhaft_2014,Sumbekova_Cartellier_Aliseda_Bourgoin_2017,Ferran_Machicoane_Aliseda_Obligado_2023,Hassaini_Coletti_2022,Hassaini_Petersen_Coletti_2023}, whereas fewer works have addressed suspensions of particles larger than the Kolmogorov scale ($D/\eta \gg 1$). \cite{Tsuji_Morikawa_1982} investigated the influence of particle size on turbulence modulation in gas-solid suspensions. They considered a horizontal pipe of diameter $30,\si{\milli\meter}$ with mean velocity between $6$ and $20,\si{\meter\per\second}$, and reported turbulence attenuation and enhancement in the presence of relatively small ($D=0.2,\si{\milli\meter}$) and large ($D=3.4,\si{\milli\meter}$) particles, respectively. \cite{Brown_Warhaft_Voth_2009} studied von K\'arm\'an turbulence generated between counter-rotating disks at Taylor-Reynolds numbers $400<\Rey_\lambda<815$. The flow was laden with neutrally buoyant polystyrene spheres whose size spanned from the dissipative range to the inertial subrange ($0.4<D/\eta<27$). They reported that the acceleration variance of the largest particles ($D/\eta>5$) follows the scaling $D^{-2/3}$ proposed by \cite{Voth_La-Porta_Crawford_Alexander_Bodenschatz_2002}. \cite{Fiabane_Zimmermann_Volk_Pinton_Bourgoin_2012} investigated homogeneous isotropic turbulence laden with particles of size $4.5<D/\eta<17$ in an icosahedral vessel. Using Voronoi tessellation analysis, they showed that neutrally buoyant particles do not cluster, whereas heavy particles cluster independently of their size. More recently, \cite{Hoque_Mitra_Sathe_Joshi_Evans_2016} examined the effect of isolated glass particles in the range $10<D/\eta<77$ in turbulence generated by an oscillating grid. They found that turbulence modulation depends critically on the ratio between the particle diameter and the integral length scale \citep{Gore_Crowe_1989}. Among the experiments on Kolmogorov-size particles, we recall \cite{Hwang_Eaton_2006_gravity}, who investigated the sedimentation of glass spheres with $\Psi_p \cong 2500$, \cite{Poelma_Westerweel_Ooms_2007}, who examined the effect of heavy particles on the decay of grid-generated turbulence, and \cite{Tanaka_Eaton_2010}, who performed high-resolution particle image velocimetry. They concluded that particles with a smaller diameter and a higher density induce the strongest attenuation in steady-state homogeneous and isotropic turbulence. Laboratory experiments and point-particle DNS of turbulent flows laden with Kolmogorov-scale solid particles and bubbles at $\Rey_\lambda \leq 180$ were compared by \cite{Volk_Calzavarini_Verhille_Lohse_Mordant_Pinton_Toschi_2008} and \cite{Tagawa_Martinez-Mercado_Prakash_Calzavarini_Sun_Lohse_2012}. The latter, who employed the datasets of \cite{Martinez-Mercado_Prakash_Tagawa_Sun_Lohse_2012}, reported stronger clustering at higher $\Rey_\lambda$. 

Gravity induces settling or rising of the dispersed phase in non-neutrally buoyant suspensions, leading to a departure from isotropy and altering both the collective dynamics of the particles and the statistics of the carrier flow substantially \citep{Brandt_Coletti_2022}. Because laboratory studies in microgravity remain extremely challenging \citep{Fallon_Rogers_2002,Hwang_Eaton_2006_gravity}, terrestrial measurements are compared with numerical simulations of homogeneous isotropic turbulence that neglect gravity \citep{Tagawa_Martinez-Mercado_Prakash_Calzavarini_Sun_Lohse_2012}. Particle-resolved numerical simulations provide a means to investigate isotropic flows in the absence of gravity, but are limited by the high spatial resolution required to accurately resolve the fluid field at sub-particle scales. Resolving the sub-particle motion via PR-DNSs requires grid spacings well below $\eta$, thus rendering the computational cost prohibitive even at moderate Reynolds numbers. For this reason, a broad body of numerical work has focused on the limits $D\ll\eta$ and $D\gg1$, while Kolmogorov-size spheres have received far less attention. Using the IBM proposed by \cite{Uhlmann_2005}, \cite{Uhlmann_Chouippe_2017} were able to investigate a particle laden triperiodic flow at $\Rey_\lambda\cong115$ and with $D/\eta=5$, slightly above the Kolmogorov scale. \cite{Schneiders_Meinke_Schröder_2017} conducted PR-DNSs of a triperiodic turbulent flow laden Kolmogorov-size spheres with $D/\eta\cong1.32$ using a Cartesian cut-cell method. They considered a suspension of $45\,000$ particles in a decaying homogeneous and isotropic turbulent flow and density ratios $\Psi_p$ in the range $40-5000$. The enormous computational burden entailed by the resolution of sub-Kolmogorov motions limited the initial value of the Taylor-based Reynolds number to $70$. \cite{Chiarini_Tandurella_Rosti_2025} used the IBM of \cite{Hori_Takagi_Rosti_2022} to characterise the behavior of a suspension of particles with diameter $D\cong\eta$ and $\Psi_p=5$, $100$. They compared their results with those obtained from point-particle simulations and reported significant discrepancies for the cases with the highest density ratio. 

\subsection{Objectives and outline}
Particle-resolved DNS of Kolmogorov-scale spheres in homogeneous isotropic turbulence have either considered decaying carrier flows without a fully developed inertial subrange at small $\Rey_\lambda$ \citep{Schneiders_Meinke_Schröder_2017}, or have been restricted to relatively low density ratios $\Psi_p \leq 100$ \citep{Uhlmann_Chouippe_2017,Chiarini_Tandurella_Rosti_2025}. However, many natural and industrial flows exhibit a fully developed inertial subrange and involve dense suspensions of high-inertia particles $\Psi_p=O(10^3)$ that can fundamentally alter the carrier-flow dynamics \citep{Jenny_Roekaerts_Beishuizen_2012,Rose_Durant_2009}.

The primary objective of this work is to quantify the impact of suspensions of heavy Kolmogorov-size spheres on the turbulent cascade, the scale-by-scale energy transfer, and the small-scale intermittency of the carrier flow, as well as to assess how increasing particle inertia shapes the collective dynamics of the dispersed phase and the structure of particle clusters. To this end, we perform large-scale particle-resolved direct numerical simulations of a homogeneous, isotropic, triply periodic turbulent flow. A Taylor-scale Reynolds number of $\Rey_\lambda\cong150$ is selected to ensure adequate scale separation within the inertial subrange in the unladen case. We consider a fixed particle volume fraction $\Phi_p=10^{-3}$ and density ratios up to $\Psi_p=1500$, a regime that lies beyond the range of validity of point-particle approaches and is directly relevant to natural and industrial flows \citep{Brandt_Coletti_2022,Marchioli_Bourgoin_Coletti_Fox_Magnaudet_Reeks_Simonin_Sommerfeld_Toschi_Wang_Balachandar_2025}.

The paper is organised as follows. The numerical method and its implementation are described in \S\ref{sec:method}, and the post-processing procedures and results are presented in \S\ref{sec:results_main}. In particular, the bulk statistics of the carrier flow are discussed in \S\ref{sec:bulk_statistics_carrier_flow}, followed by energy spectra and structure functions in \S\ref{sec:energy_spectra_structure_functions}, the scale-by-scale energy budget in \S\ref{sec:scale-by-scale}, and the analysis of the velocity-gradient tensor in \S\ref{sec:velocity_gradient_tensor}. Particle dynamics and clustering statistics are examined in \S\ref{sec:particle_motion} and \S\ref{sec:particle_clustering}, respectively, while the particles' preferential sampling is discussed in \ref{sec:preferential_sampling}. Finally, conclusions are drawn in \S\ref{sec:conclusions}.

\section{Direct numerical simulations}\label{sec:method}
\subsection{Mathematical formulation}
We consider a triply periodic cubic domain of volume $\mathcal{V}_L=L^3$, where $L$ is the side length of the cube. The carrier flow (fluid phase) is governed by the incompressible Navier-Stokes equations for momentum balance and the continuity constraint \par\vspace{-\baselineskip}%
\begin{subequations} \label{eq:fluid_governing}
\begin{equation}  
  \frac{\p u_i}{\p t} + \frac{\p u_j u_i}{\p x_j} = \frac{1}{\rho_f}\frac{\p\sigma_{ij}}{\p x_j} + f_i^\ABC + f_i^{\particle}, \label{eq:NS_momentum}
\end{equation}
\begin{equation}
  \frac{\p u_k}{\p x_k} = 0, \label{eq:NS_mass} 
\end{equation}
\end{subequations}
where Einstein summation over repeated indices is implied. Here, $u_i$ is the fluid velocity, $\rho_f$ the fluid density, and $f_i^{\particle}$ is the IBM forcing exerted by the solid particles on the fluid \citep{Hori_Takagi_Rosti_2022}. The large-scale volume forcing $f_i^\ABC$ is of the Arnold-Beltrami-Childress (ABC) type \citep[p. 20]{Arnold_1965} \par\vspace{-\baselineskip}%
\begin{subequations}\label{eq:ABC_forcing}
\begin{align}
    f_x^\ABC &= F_0 \left[ \cos\left(\frac{2\pi}{L}y\right) + \sin\left(\frac{2\pi}{L}z\right) \right], \\
    f_y^\ABC &= F_0 \left[ \cos\left(\frac{2\pi}{L}z\right) + \sin\left(\frac{2\pi}{L}x\right) \right], \\
    f_z^\ABC &= F_0 \left[ \cos\left(\frac{2\pi}{L}x\right) + \sin\left(\frac{2\pi}{L}y\right) \right],
\end{align}
\end{subequations}
and has an amplitude $F_0$ with units of force per mass. Because the carrier flow is Newtonian and incompressible, the stress tensor $\sigma_{ij}$ reads \par\vspace{-\baselineskip}%
\begin{equation}\label{eq:cauchy_stress_tensor}
    \sigma_{ij} = - p \delta_{ij} + 2\mu_f S_{ij},
\end{equation}
where $p$ is the fluid pressure, $\delta_{ij}$ is the Kronecker delta, $\mu_f$ is the dynamic viscosity and $S_{ij} = (\p_ju_i + \p_iu_j)/2$ is the symmetric part of the velocity gradient tensor $\p_ju_i$. From this point onwards, the kinematic viscosity of the fluid phase $\nu_f=\mu_f/\rho_f$ will be used.

The velocity of a point inside a solid spherical particle is due to the contribution of the translational velocity and the rotation around the particle center $U_i = \tilde U_{i} + \levicivita \Omega_{j} R_k$, where $\Omega_j$ is the particle rotational velocity, $\levicivita $ the Levi-Civita tensor, and $R_k$ the radial distance from the particle center. The time evolution of the translational velocity $\tilde{U}_i$ and the angular velocity $\Omega_j$ is governed by the Newton-Euler equations\par\vspace{-\baselineskip}%
\begin{subequations}\label{eq:newton_euler}
\begin{align}
    m_p \frac{\d \tilde U_i}{\d t} &= \oint_{\p \mathcal{V}_p} \sigma_{ij} n_j \d\left(\p \mathcal{V}\right) + F_{i}^{\collision}, \label{eq:newton_euler_momentum}\\
    \mathcal{I}_p \frac{\d \Omega_i}{\d t} &= \oint_{\p \mathcal{V}_p} \levicivita R_j \sigma_{kl} n_l \d\left(\p \mathcal{V}\right).
\end{align}
\end{subequations}
Here, $F^{\collision}$ is the force exerted by inter-particle collisions, $\sigma_{ij}$ is defined in \eqref{eq:cauchy_stress_tensor}, $n_i$ is the outer normal to the particle surface, $m_p = (\pi/6) \rho_p D^3$ and $\mathcal{I}_p = (m_p/10) D^2$ are the particle mass and moment of inertia, respectively, and $D$ is the particle diameter. No-slip and no-penetration conditions are applied at the particle surface $u_i = U_i$, where the velocity of the fluid and solid phase match. The particle translational velocity is integrated at every time step to update the particle location $X_i$, where $\tilde U_i = \d X_i/\d t$. Inter-particle collisions are handled with the mass-spring-dash pot model of \cite{Tsuji_Kawaguchi_Tanaka_1993} and \cite{Costa_Boersma_Westerweel_Breugem_2015}. A repulsive force acts on the particles as soon as their volumes overlap, i.e. when the distance of their centres falls below $D$. 

\begin{table}
\centering
\begin{tabular*}{0.75\textwidth}{@{\extracolsep\fill}ccccc}
$N$ & $n_{\text{grid}}$ & $\Phi_p$ & $D/L$ & $\Rey_L$ \\
$74\,208$ & $2048^3$ & $10^{-3}$ & $0.003$ & $1761$ 
\end{tabular*}
\caption{Parameters common to all simulations: $N$ the total number of particles, $n_{\text{grid}}$ the number of grid points in the uniform cartesian mesh, $\Phi_p$ the particle volume fraction, $D$ the particle diameter, and $\Rey_L\equiv\left(F_0L^3\right)^{1/2}/\nu_f$ the Reynolds number defined on the characteristic velocity $\left(F_0L\right)^{1/2}$. Here, $F_0=5$ is the amplitude of the ABC volume forcing \eqref{eq:ABC_forcing} and has units of force per mass, $L=2\pi$ is the side length of the cubic domain and $\nu_f=\mu_f/\rho_f=1/50.$} \label{tab:simulation_parameters}
\end{table}

The parameters common to all simulations are summarised in table \ref{tab:simulation_parameters}. The Reynolds number based on the side length of the domain and the characterstic velocity $(F_0L)^{1/2}$ is $\Rey_L\equiv F_0^{1/2}L^{3/2}/\nu_f = 1761$. We consider a monodisperse (i.e. uniform in size) population of $N=74\,208$ spherical particles with diameter $D \cong 0.003L$ that matches the Kolmogorov scale $\eta$ in all the cases examined. The particle diameter is kept constant and chosen to match the space-time average of the Kolmogorov scale in the unladen case ($M_p=0$). As verified \textit{a posteriori}, moderate variations in the dissipation rate with $M_p$ lead to negligible changes in $\eta$ (see discussion in \S\ref{sec:bulk_statistics_carrier_flow}). The volume fraction of the solid phase is kept constant $\Phi_p = (\pi N/6) (D/L)^3 = 10^{-3}$ across all simulations. Thus, the mass fraction is determined solely by the density ratio $\Psi_p= \rho_p/\rho_f$ through the relation\par\vspace{-\baselineskip}%
\begin{equation}\label{eq:mass_fraction_def}
  M_p = \frac{\Psi_p\Phi_p}{1 - \Phi_p + \Psi_p\Phi_p}.
\end{equation}
Four laden cases with density ratio $\Psi_p = 100$, $250$, $666$, $1499$ and mass fraction $M_p=0.1$, $0.2$, $0.4$, $0.6$ are compared to the unladen case ($\Phi_p=M_p=0$). These values of $M_p$ match those used in the experiments of \cite{Tanaka_Eaton_2010} (see table 3 therein).

\subsection{Numerical method}
The numerical method described in this section is implemented on the in-house Fortran code \href{https://www.oist.jp/research/research-units/cffu/fujin}{\textit{Fujin}} \citep{Rosti_2026}. The momentum equation for the fluid phase \eqref{eq:NS_momentum} is advanced in time with a fractional step method \citep{Amsden_Harlow_1970,Kim_Moin_Moser_1987,Tome_Duffy_McKee_1996}. The nonlinear and viscous terms in \eqref{eq:NS_momentum} are discretised using an Adams-Bashforth scheme in the predictor step, while incompressibility is enforced by solving the Poisson equation for the pressure with a fast Fourier transform. The interplay of the fluid phase and the solid spheres is taken into account with the Eulerian immersed boundary method (IBM) described by \cite{Hori_Takagi_Rosti_2022}. A fixed-radius, near-neighbours search algorithm is used to reduce the computational cost of the collisions from $O(N^2)$ to $O(N)$ \citep{Monti_Rathee_Shen_Rosti_2021}. The code uses a two-dimensional Cartesian parallelisation provided by the Message-Passage Interface (MPI) library. The triperiodic cubic domain is discretised using an uniform cartesian grid of $n_{\text{grid}} = 2048^3$ points. The computations were performed using 4096 cores on the CPU-based cluster \href{https://www.oist.jp/equipment/scda-deigo}{Deigo} at the Okinawa Institute for Science and Technology. 

The laden case with mass fraction $M_p=0.1$ was initialised by seeding a fully developed turbulent flow with $N$ spherical particles. The procedure is the following. The coordinates $X_i$ of each particle are sampled sequentially from a uniform distribution using a random-number generator in Matlab. If a generated particle does not overlap with any previously placed particle, it is accepted; otherwise, it is rejected and the procedure is repeated until all initial positions are assigned. The particles are initially at rest ($\tilde U_i= \Omega_i=0$) and subsequently accelerate under the hydrodynamic forcing exerted through the surface integrals in \eqref{eq:newton_euler}. The introduction of the particles induces a transient phase during which the carrier flow evolves toward a new statistically stationary state. To reduce the duration of this transient and to limit the associated computational cost, the cases with higher mass fractions $M_p=0.2$, $0.4$ and $0.6$ are instead initialised from a fully developed particle-laden turbulent flow at lower mass fraction by progressively increasing the density ratio $\Psi_p$.

To assess the statistical convergence of carrier-phase quantities such as the kinetic energy $K(x_i,t)=u_iu_i/2$ and the dissipation rate $\varepsilon(x_i,t) = 2\nu_fS_{ij}S_{ij}$, we computed their averages over the entire domain. These volume averages are denoted by $\averagespace{\cdot} = L^{-3}\int_{\mathcal{V}}\left(\cdot\right)\d\mathcal{V}$. In addition, temporal averages are defined as $\averagetime{\cdot} = T_{\text{tot}}^{-1}\int_{T}\left(\cdot\right)\d t$, evaluated over the full duration of the simulation after discarding the initial transient $T_{\text{tot}}$. A fixed time step $\Delta t = 2\times10^{-5}$ ($\Delta t/\tau_\eta\cong10^{-3}$) was employed for all cases. Upon the introduction of the particles, approximately $40\,000$ time steps were required to reach statistical convergence for the case $M_p=0.1$ and $25\,000$ for the cases $M_p=0.2$, $0.4$ and $0.6$. After discarding this initial transient, all simulations were advanced for approximately $30\averagetime{\tau_{\mathcal{L}}}$, where $\tau_{\mathcal{L}}(t) = \mathcal{L}/\urms$ denotes the large-scale eddy turnover time. This quantity is defined in terms of the integral length scale $\mathcal{L}(t)$ and the root-mean-square velocity $\urms(t) = \left(2\averagespace{K}/3\right)^{1/2}$. For homogeneous and isotropic turbulence, the integral length scale is given by \citep[\S 6.5]{Pope_2000} \par\vspace{-\baselineskip}%
\begin{equation}\label{eq:integral_scale}
  \mathcal{L}\left(t\right) = \frac{3\pi}{4}\left[\int_0^\infty \kappa^{-1}\hat E(\kappa,t)\d\kappa\right]\left[\int_0^\infty \hat E(\kappa,t)\d\kappa\right]^{-1},
\end{equation} 
where $\hat E(\kappa,t)$ is the energy spectrum. The latter is defined such that $\int_0^{\infty}\hat E(\kappa,t)\d\kappa = \averagespace{K}(t)$, i.e. it recovers the instantaneous kinetic energy of the carrier flow.
\section{Postprocessing and results}\label{sec:results_main}
We present and discuss the results of the PR-DNSs performed using the method described in \S\ref{sec:method}. Since the volume fraction is fixed, each density ratio $\Psi_p$ corresponds to a unique mass fraction $M_p$ through \eqref{eq:mass_fraction_def}. For clarity, the laden cases are therefore identified by $M_p$. In the present setup, the Stokes number (i.e.\ the ratio of the particle response time to a characteristic flow time scale) covaries with $M_p$ and  $\Psi_p$, so that their individual effects cannot be disentangled. As shown in \S\ref{sec:particle_motion}, an accurate evaluation of the particle response time (and hence of the Stokes number) requires nonlinear corrections, implying that the Stokes number is not a single well-defined quantity but rather a distributed variable \citep{Balachandar_Eaton_2010}, with PDFs that broaden as the mass loading increases. For these reasons, we present the results as a function of $M_p$, which is unique for each case.

\subsection{Bulk statistics of the carrier flow}\label{sec:bulk_statistics_carrier_flow}
\begin{figure}
  \includegraphics[width=\linewidth]{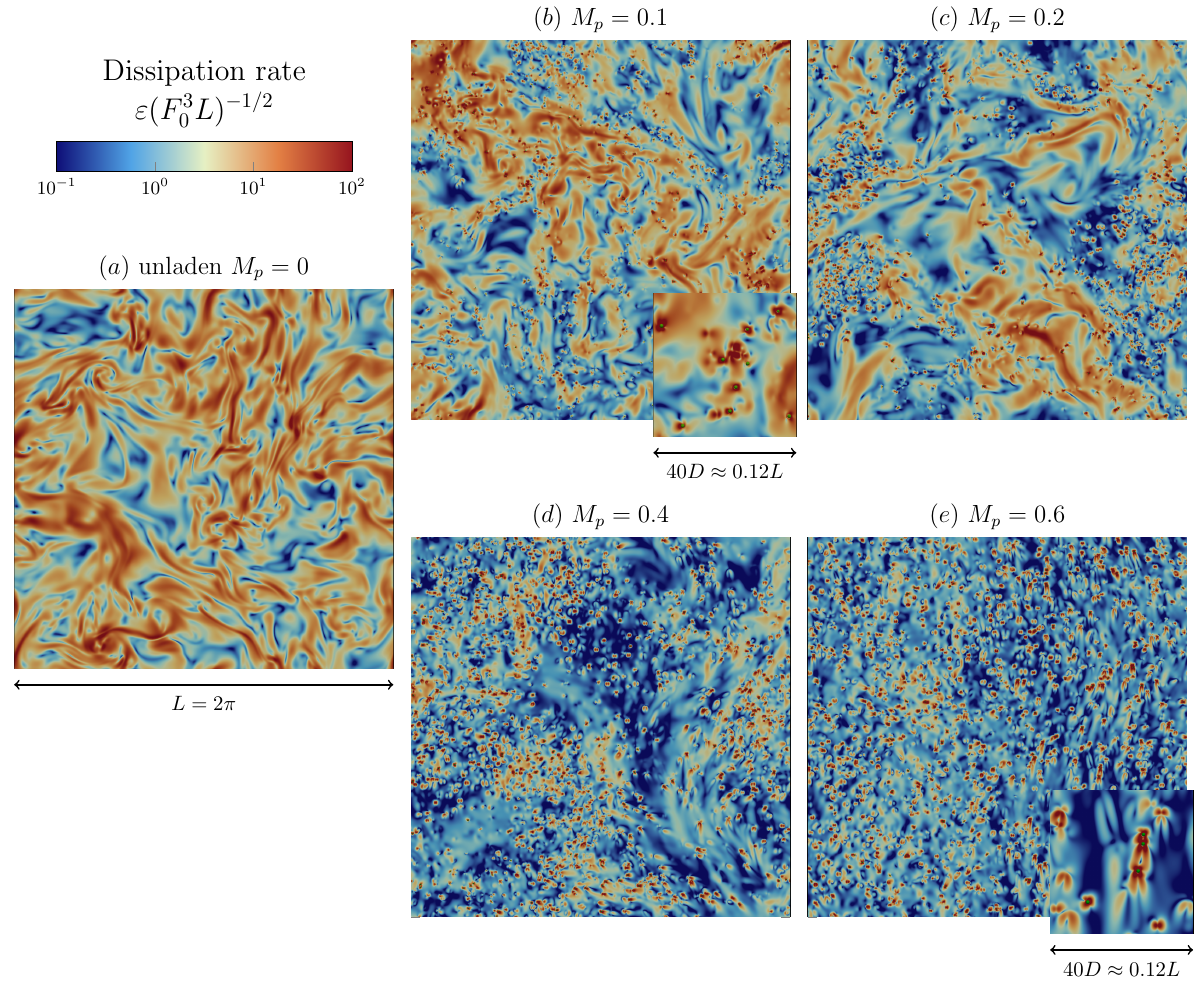}
  \caption{Instantaneous visualizations of the non-dimensional turbulent dissipation rate, $\varepsilon (F_0^3L)^{-1/2}$, on a planar section of size $L^2=(2\pi)^2$. The dissipation field is displayed in each panel using a logarithmic colour scale spanning the range $[10^{-1},10^2]$. The leftmost panel (\textit{a}) corresponds to the unladen, single-phase flow, whereas the four panels on the right depict the particle-laden cases with mass fractions $M_p = 0.1$ (\textit{b}), $0.2$ (\textit{c}), $0.4$ (\textit{d}), and $0.6$ (\textit{e}). Insets in panels (\textit{b}) and (\textit{e}) show magnified views of the dissipation field in the vicinity of particles (green spheres), over a subdomain of size $40D \approx 0.12L$.}
  \label{fig:visualization_dissipation_rate}
\end{figure}

\begin{figure}
  \includegraphics[width=\linewidth]{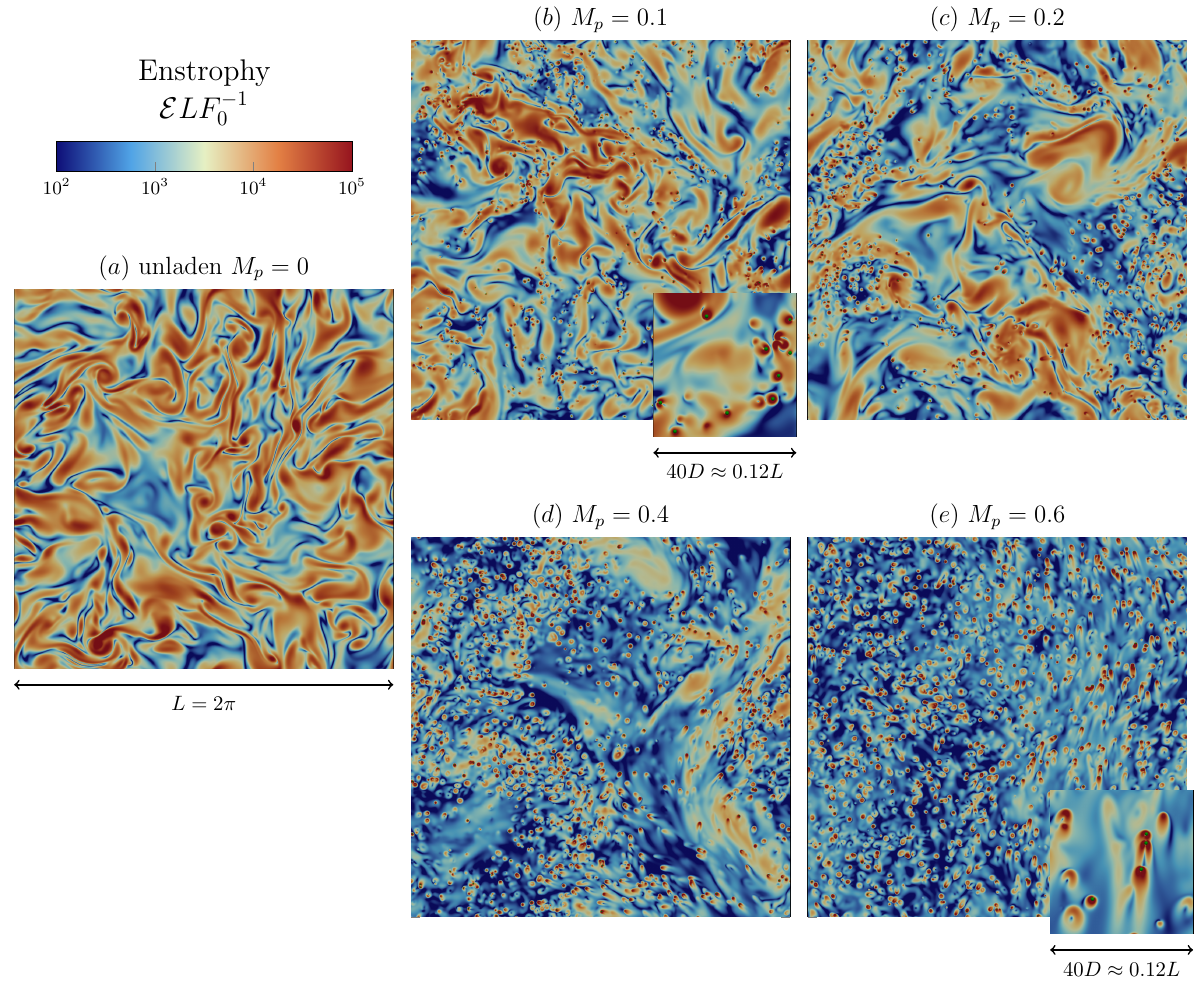}
  \caption{Instantaneous visualizations of the non-dimensional turbulent enstrophy, $\mathcal{E} LF_0^{-1}$, on a planar section of size $L^2=(2\pi)^2$. The enstrophy field is displayed in each panel using a logarithmic colour scale spanning the range $[10^{2},10^5]$. The leftmost panel (\textit{a}) corresponds to the unladen, single-phase flow, whereas the four panels on the right depict the particle-laden cases with mass fractions $M_p = 0.1$ (\textit{b}), $0.2$ (\textit{c}), $0.4$ (\textit{d}), and $0.6$ (\textit{e}). Insets in panels (\textit{b}) and (\textit{e}) show magnified views of the enstrophy field in the vicinity of particles (green spheres), over a subdomain of size $40D \approx 0.12L$.}
  \label{fig:visualization_enstrophy}
\end{figure}

Increasing the inertia of Kolmogorov-size particles has a dramatic effect on the large-scale dynamics of the turbulent carrier flow. The five panels in figure \ref{fig:visualization_dissipation_rate} show the effect of $M_p$ (and $\Psi_p$) on the dissipation rate $\varepsilon = 2\nu_fS_{ij}S_{ij}$, where $S_{ij}$ is the symmetric part of the velocity gradient tensor defined in \eqref{eq:cauchy_stress_tensor}. The values of $\varepsilon$ are normalised on the reference quantity $(F_0^3L)^{1/2}$, which has units of energy per mass per time. The colour scale is logarithmic and ranges between $10^{-1}$ (dark blue) and $10^2$ (dark orange), with the particles' location marked by the localised spots of high dissipation associated to the shear layers at the particles' surface \citep{Cannon_Olivieri_Rosti_2024}. All the cases show large excursions in $\varepsilon$. The left panel of figure \ref{fig:visualization_dissipation_rate}$a$ shows the unladen flow field, characterised by large regions of high dissipation (yellow and orange) and low dissipation (dark blue). Introducing particles at the lowest mass fraction ($M_p=0.1$, figure \ref{fig:visualization_dissipation_rate}$b$) does not alter these structures significantly: particles preferentially reside in high-$\varepsilon$ regions while avoiding low-$\varepsilon$ ones. These structures break down into smaller ones and the peaks of large $\varepsilon$ become confined to the particles' shear layers as $M_p$ increases to 0.2 (\ref{fig:visualization_dissipation_rate}$c$) and 0.4 (\ref{fig:visualization_dissipation_rate}$d$). When $M_p=0.6$ (\ref{fig:visualization_dissipation_rate}$e$), the large regions of high $\varepsilon$ vanish: the flow field is characterised by smaller, adjacent regions of moderate (yellow) and low (blue) $\varepsilon$, while the dissipation remains still elevated in the particles' wakes. As can be observed in figure \ref{fig:visualization_enstrophy}, increasing the particle inertia has a similar effect on the enstrophy $\mathcal{E} = \omega_i\omega_i/2$, where $\omega_i$ is the vorticity. Similarly to the dissipation, these large regions of intense vorticity disappear as $M_p$ increases, and high values of the enstrophy remain confined to the particles' wakes. As shown in the insets of figures \ref{fig:visualization_dissipation_rate}$b$, $e$ and of figures \ref{fig:visualization_enstrophy}$b$, $e$, elongated laminar wakes appear as $M_p$ increases and the carrier flow becomes more intermittent at the particle scale. 

\begin{table}
\centering
\begin{tabular*}{\textwidth}{@{\extracolsep\fill}ccccccccccc}
& $\Psi_p$ & $M_p$ & $\displaystyle\frac{\averagespacetime{K}}{F_0L}$ & $\displaystyle\frac{\averagespacetime{\varepsilon}}{(F_0^3L)^{1/2}}$ & $\displaystyle\frac{\averagetime{\eta}}{L}$ & $\displaystyle \frac{\averagetime{\tau_\eta}}{\left(L/F_0\right)^{1/2}}$ & $\displaystyle \frac{\averagetime{\lambda}}{L}$ & $\averagetime{\Rey_\lambda}$ & $\displaystyle \frac{\averagetime{\mathcal{L}}}{L}$ & $\displaystyle \frac{\averagetime{\tau_{\mathcal{L}}}}{\left(L/F_0\right)^{1/2}}$ \\
\midrule
\tikz[baseline=-0.5ex]\draw[fill=black, draw=none] (-0.3em,-0.3em) rectangle (1.8em,0.6em); & -      & $0$   & $2.11$ & $2.40$ & $0.0030$ & $0.016$ & $0.07$ & $149.34$ & $0.27$ & $0.25$ \\
\tikz[baseline=-0.5ex]\draw[fill=mp01, draw=none] (-0.3em,-0.3em) rectangle (1.8em,0.6em); & $100$  & $0.1$ & $1.60$ & $1.89$ & $0.0032$ & $0.018$ & $0.07$ & $127.29$ & $0.27$ & $0.28$ \\
\tikz[baseline=-0.5ex]\draw[fill=mp02, draw=none] (-0.3em,-0.3em) rectangle (1.8em,0.6em); & $250$  & $0.2$ & $1.47$ & $1.83$ & $0.0032$ & $0.018$ & $0.07$ & $117.94$ & $0.29$ & $0.33$ \\
\tikz[baseline=-0.5ex]\draw[fill=mp04, draw=none] (-0.3em,-0.3em) rectangle (1.8em,0.6em); & $666$  & $0.4$ & $1.42$ & $2.03$ & $0.0031$ & $0.017$ & $0.06$ & $108.57$ & $0.33$ & $0.37$ \\
\tikz[baseline=-0.5ex]\draw[fill=mp06, draw=none] (-0.3em,-0.3em) rectangle (1.8em,0.6em); & $1499$ & $0.6$ & $1.42$ & $2.07$ & $0.0031$ & $0.017$ & $0.06$ & $107.58$ & $0.37$ & $0.40$ \\
\end{tabular*}
\caption{Statistics of the carrier flow. The first row describes the unladen case and the others the laden cases. The colours correspond to the cases shown in figures \ref{fig:energy_spectra}, \ref{fig:structure_functions}, \ref{fig:skewness_kurtosis}, \ref{fig:scalebyscale_energy_transfer} and \ref{fig:Q_and_R_PDF}. The temporal averages are denoted by the operator $\averagetime{\cdot}$ and the spatial-temporal averages by the operator $\averagespacetime{\cdot}$. The mass fraction of the solid phase $M_p$ is defined in \eqref{eq:mass_fraction_def}. All the quantities are non-dimensional. The averages of the kinetic energy $\averagespacetime{K}$, the dissipation rate $\averagespacetime{\varepsilon}$, the Kolmogorov spatial scale $\averagetime{\eta}$ and temporal scale $\averagetime{\tau_\eta}$, the Taylor length scale $\averagetime{\lambda}$, the Taylor-based Reynolds number $\averagetime{\Rey_\lambda}$, the integral scale $\averagetime{\mathcal{L}}$ \eqref{eq:integral_scale} and the eddy turnover time $\averagetime{\tau_\mathcal{L}}$ are listed.\label{tab:flow_statistics}}
\end{table}

Figures \ref{fig:visualization_dissipation_rate} and \ref{fig:visualization_enstrophy} suggest that the broad range of spatial scales and the direct energy cascade typical of homogeneous and isotropic turbulent flows disappear as the density of the particles increases. To gain further insights on the effect of the particle inertia on the physical properties of the carrier flow, which are point-wise quantities local in space and time, we compute their spatial averages over the domain $\averagespace{\cdot}$, their temporal averages $\averagetime{\cdot}$ and their space-time averages $\averagespacetime{\cdot} = \averagetime{\averagespace{\cdot}}$. Their values are made non-dimensional and listed in table \ref{tab:flow_statistics} where the four laden cases shown in the visualizations \ref{fig:visualization_dissipation_rate} and \ref{fig:visualization_enstrophy} are compared with the unladen case $M_p=0$. From this point onwards, we shall denote the results of the laden cases with violet ($M_p=0.1$), teal ($M_p=0.2$), green ($M_p=0.4$) and yellow ($M_p=0.6$) curves. The kinetic energy is $K = u_iu_i/2$, the dissipation rate was defined earlier in this section as $\varepsilon = 2\nu_fS_{ij}S_{ij}$, the Kolmogorov length scale is $\eta(t) = (\nu_f^3/\averagespace{\varepsilon})^{1/4}$ the Kolmogorov time scale is $\tau_\eta(t) = (\nu_f/\averagespace{\varepsilon})^{1/2}$, the transverse Taylor length scale is $\lambda(t) = (10\nu_f\averagespace{K}/\averagespace{\varepsilon})^{1/2}$ and the Taylor-based Reynolds number is $\Rey_\lambda(t) = \urms\lambda/\nu_f$ \citep[\S 6.5]{Pope_2000}. 

As reported in table \ref{tab:flow_statistics}, the average kinetic energy $\averagespacetime{K}$, the average dissipation $\averagespacetime{\varepsilon}$ and the Taylor-Reynolds number $\averagetime{\Rey_\lambda}$ decrease markedly as $M_p$ increases: a significant portion of the turbulent kinetic energy is transferred directly to the particles via the fluid-particle forcing $f_i^{\particle}$ in \eqref{eq:NS_momentum} and the surface integrals in \eqref{eq:newton_euler}. The average integral length scale $\averagetime{\mathcal{L}}$ \eqref{eq:integral_scale} and the average integral time scale $\averagetime{\tau_{\mathcal{L}}}$ of the flow increase slightly with particle inertia, suggesting that the energy density shifts to lower wavenumbers while the intermediate and small-scale structures weaken as $M_p$ increases. As shown in \S\ref{sec:energy_spectra_structure_functions}, the monotonic increase of the integral length scale should be interpreted as a redistribution of kinetic energy across scales, rather than as evidence of larger turbulent structures. Overall, the amount of kinetic energy lost by the carrier flow increases when suspensions of larger mass fraction are considered. In the laden cases, $\averagespacetime{K}$ decreases as $M_p$ increases from $0.1$ to $0.4$, while $\averagespacetime{\varepsilon}$ increases slightly from $M_p=0.2$ to $M_p=0.4$. The Kolmogorov spatial scale $\eta$ and the temporal scale $\tau_\eta$ are not noticeably affected by $M_p$ in any of the cases listed in table \ref{tab:flow_statistics}, thus confirming that $D\cong\eta$. In particular, the dependence of $\eta$ on the quartic root of the dissipation rate implies a weak sensitivity to the latter's variations: values of $\averagetime{\eta}$ differ by at most $7\%$. These findings are in good agreement with the laboratory measurements of \cite{Tanaka_Eaton_2010}, who reported an attenuation of 25\% of the average turbulent kinetic energy when seeding a single-phase homogeneous and isotropic turbulent flow at $\Rey_\lambda\cong130$ with Kolmogorov-size particles with $D/\eta\cong2$, $M_p=0.4$ and $\Phi_p=1.8\times10^{-4}$ (see figure 16 therein). In the present DNSs we observe a $32\%$ decrease of $\averagespacetime{K}$ when the unladen turbulent flow at $\Rey_p\cong150$ is seeded with a suspension of volume fraction $\Phi_p=10^{-3}$ and mass fraction $M_p=0.4$. 

\subsection{Kinetic energy spectra and structure functions}\label{sec:energy_spectra_structure_functions}
\begin{figure}
  \centering
  \includegraphics[width=\linewidth]{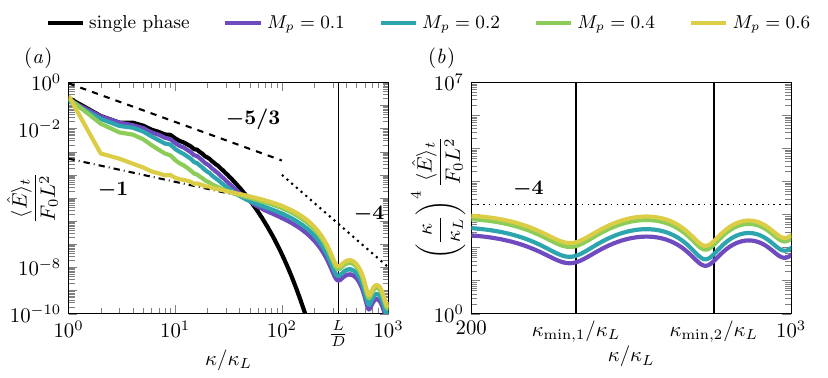}
  \caption{($a$) Kinetic energy spectra $\averagetime{\hat E}$ and ($b$) compensated energy spectra $(\kappa/\kappa_L)^4\averagetime{\hat E}$, for different mass fraction $M_p$ of the suspension. All wavenumbers are scaled by $\kappa_L=2\pi/L$ and the spectra by the reference quantity $F_0L^2$. The latter has units of energy per wavenumber per unit mass. The colours correspond to the cases listed in table \ref{tab:flow_statistics}. The black dashed line, the dash-dot line, and the black dotted line represent the $(\kappa/\kappa_L)^{-5/3}$, $(\kappa/\kappa_L)^{-1}$, and $(\kappa/\kappa_L)^{-4}$ scaling, respectively. The black solid vertical line in ($a$) denotes the wavenumber of the particle diameter $\kappa_D/\kappa_L = L/D$, while the vertical lines in ($b$) represent the minima of the high-wavenumber oscillations $\kappa_{min,m} = m\pi/D$, $m\in\mathbb{N}$.}
  \label{fig:energy_spectra}
\end{figure}

To understand how the particles affect the distribution of turbulent kinetic energy across the scales of motion, we compare the kinetic energy spectra and the longitudinal structure functions computed for different mass fractions $M_p$. We computed the spectra and structure functions by considering all Eulerian grid points, including those located inside the particles \citep{Wang_Ayala_Gao_Andersen_Mathews_2014,Olivieri_Cannon_Rosti_2022,Chiarini_Tandurella_Rosti_2025,Jiang_Brandt_Xu_Zhao_2025,Jiang_Mirzareza_Crialesi-Esposito_Brandt_2026}. Within this Eulerian IBM framework, the fluid velocity inside the particles is governed by the Navier-Stokes equations \eqref{eq:fluid_governing} and satisfies no-slip and no-penetration boundary conditions at the particle surface. No additional treatment -- such as enforcing zero velocity inside the particles \citep{Lucci_Ferrante_Elghobashi_2010} -- is applied. As a result, the mean and the fluctuating fluid velocity inside the particles are comparable in magnitude to $\tilde U_i$. As verified by \cite{Chiarini_Rosti_2024} and \cite{Chiarini_Tandurella_Rosti_2025}, the two-point structure function is not affected by the inclusion of Eulerian points inside the particles for solid volume fractions $\Phi_p\leq10^{-2}$. The spectra are shown in figure \ref{fig:energy_spectra}, where all the spatial wavenumbers $\kappa$ are scaled with respect to the wavenumber of the domain $\kappa_L = 2\pi/L$. Each curve in figure \ref{fig:energy_spectra}$a$ describes the effect of a different mass fraction $M_p$. While the spectra of the unladen case (black curve, $\averagetime{\Rey_\lambda}\cong 150$) and the case $M_p=0.1$ (red curve, $\averagetime{\Rey_\lambda}\cong127$) feature the classical Kolmogorov scaling $(\kappa/\kappa_L)^{-5/3}$ for $\kappa/\kappa_L<20$, the spectra obtained for larger $M_p$ depart from such scaling gradually as particle inertia increases. When compared to the unladen case, all the spectra show a lower energy density for $\kappa/\kappa_L<20$ and a higher energy density at wavenumbers in the dissipative range $\kappa/\kappa_L\geq10^2$, meaning that Kolmogorov-size particles weaken the large-scale structures of the flow and energise small-scale motions at wavenumbers comparable-to or larger-than the particle wavenumber $\kappa_D = 2\pi/D$ (solid vertical line in figure \ref{fig:energy_spectra}$a$). The amount of kinetic energy transferred to the smallest scales of motion ($\kappa/\kappa_L>10^2$) increases slightly with $M_p$. This behavior was reported previously in decaying isotropic turbulence \citep{Lucci_Ferrante_Elghobashi_2010,Schneiders_Meinke_Schröder_2017} and confirms the picture of figures \ref{fig:visualization_dissipation_rate} and \ref{fig:visualization_enstrophy}: high dissipation and high enstrophy concentrate in the particle wakes as the density of the dispersed phase increases. The modification of the spectra also explains the increase in the integral scale reported in table \ref{tab:flow_statistics}. As the mass loading increases, the spectrum loses relatively more energy at intermediate wavenumbers ($2<\kappa<30$) than at large scales ($\kappa = O(1)$). Since this depletion is not compensated by a sufficient increase of kinetic energy at the particle scale ($\kappa \gg 1$), the weighted integral in \eqref{eq:integral_scale} increases, leading to a larger value of $\mathcal{L}$. The instantaneous spectra $\hat E(\kappa,t)$ do not depart significantly from the temporal averages shown in figure \ref{fig:energy_spectra}, which indicates that the energy reorganization described in this section is statistically stationary.

The compensated spectra $(\kappa/\kappa_L)^4\averagetime{\hat E}$ shown in figure \ref{fig:energy_spectra}$b$ prove that, for large $\kappa$, $\averagetime{\hat E}$ follows the scaling $(\kappa/\kappa_L)^{-4}$ reported by \cite{Chiarini_Tandurella_Rosti_2025} for Kolmogorov-size spherical particles of lower density. The same scaling is observed in the turbulent spectra of \citet[see figure 6 therein]{Wang_Ayala_Gao_Andersen_Mathews_2014}, who considered a suspension of spheres with markedly different properties $\Psi_p=5$, $\Phi_p=0.1$ and $D/\eta=8$. The spectra of \citet[see figure 6 therein]{Jiang_Mirzareza_Crialesi-Esposito_Brandt_2026}, who studied Taylor-size settling spheres, show a comparable scaling for the case of a very dense suspension. A similar multiscale behaviour $(\kappa/\kappa_L)^{-3}$ has been reported in the experimental and numerical literature on bubbly flows \citep[e.g.][]{Prakash_Martinez-Mercado_van-Wijngaarden_Mancilla_Tagawa_Lohse_Sun_2016}, while laboratory measurements of turbulent flows seeded with Kolmogorov-size particles have not provided a complete characterisation of the energy spectra at high wavenumbers $O(\eta^{-1})$ \citep{Hwang_Eaton_2006_gravity,Poelma_Westerweel_Ooms_2007,Tanaka_Eaton_2010}. Recently, \cite{Ramirez_Burlot_Zamansky_Bois_Risso_2024} argued that the sub-particle power-law decay and the periodic oscillations arise from singularities at the interface between the two phases \citep{Wang_Ayala_Gao_Andersen_Mathews_2014}. Their theoretical analysis predicts (i) a power-law decay with exponent $-2q-2$, where $q$ denotes the order of the singularity, and (ii) oscillations with minima located at $\kappa_{\mathrm{min},m} = m\pi/D$, with $m \in \mathbb{N}$ \citep{Lucci_Ferrante_Elghobashi_2010}. In particular, for a velocity field that is continuous across the interface but has a discontinuous first derivative, the singularity has order $q=1$ and $E(\kappa)=O(\kappa^{-4})$ at large $\kappa$. High-wavenumber oscillations are clearly visible in figure \ref{fig:energy_spectra}$b$, and the locations of the minima are in good agreement with $\kappa_{\mathrm{min},m}$. In the present simulations, the second-order structure functions of the second velocity differences (not shown) are $\sim r^3$ for small separations $r$, confirming that the observed $\kappa^{-4}$ scaling is not a numerical artifact \citep{Biferale_Cencini_Lanotte_Vergni_2003}. However, high-resolution laboratorial measurements of the spectra in the wavenumber range $\kappa\geq\kappa_D$ are necessary to confirm the physical nature of the $(\kappa/\kappa_L)^{-4}$ scaling.

\begin{figure}
  \centering
  \includegraphics[width=\linewidth]{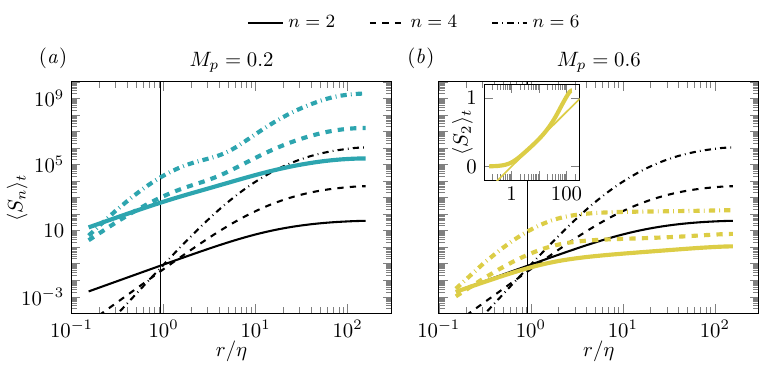}
  \caption{Temporal averages of the structure functions of the fluid velocity $S_n$ with $n=2$ (solid) $n=4$ (dashed) and $n=6$ (dash dot). The thick solid curves represent the laden cases for $M_p=0.2$ ($a$) and $M_p=0.6$ ($b$), while the structure functions of the unladen flow field are also plotted for comparison with thin curves. The thin vertical line denotes the particle diameter normalised with the Kolmogorov length $D/\eta$. The inset in ($b$) compares the semi-logarithmic plot of $S_2$ along with the plot of $\log(r/\eta)$, for $M_p=0.6$.}
  \label{fig:structure_functions}
\end{figure}

When the carrier flow is seeded with the heaviest particles (brown curve in figure \ref{fig:energy_spectra}), the kinetic energy spectrum follows the scaling $(\kappa/\kappa_L)^{-1}$ over the range $2 \leq \kappa/\kappa_L < 70$ (brown dash-dot line in figure \ref{fig:energy_spectra}$a$). The same scaling was reported by \cite{Olivieri_Cannon_Rosti_2022} in flows laden with slender fibres of Kolmogorov-scale cross section. In that study, the authors argued that heavier fibres acted as barriers between points separated by distances larger than the fibre cross section, thereby decorrelating the flow and modifying both the exponent of the second-order structure function and the scaling of the energy spectrum. To assess whether an analogous mechanism operates in the presence of heavy Kolmogorov-size particles, we compute the longitudinal structure functions of the first differences $S_n(r) = \averagespace{[\delta u(r,t)]^n}$, where $\delta u(r,t) = [\boldsymbol{u}(\boldsymbol{x} + \boldsymbol{r},t) - \boldsymbol{u}(\boldsymbol{x},t)] \cdot \boldsymbol{r}/\abs{\boldsymbol{r}}$ denotes the longitudinal velocity increment \citep[\S 6.7.2]{Pope_2000}. Figure \ref{fig:structure_functions} shows the second- ($n=2$), fourth- ($n=4$), and sixth-order ($n=6$) structure functions. The laden cases (thick solid curves) for $M_p=0.2$ (panel $a$) and $M_p=0.6$ (panel $b$) are compared with the unladen case (thin dashed curves). For $M_p=0.2$, the second-order structure function exhibits an approximately constant offset from the unladen case and a power-law range extending to separations $O(10\eta)$, consistently with the inertial subrange observed in figure \ref{fig:energy_spectra}$a$. As particle inertia increases, both the magnitude and the slope of the structure functions decrease markedly. In the heaviest-particle case, the fourth- and sixth-order structure functions approach a plateau and their growth saturates at scales much smaller than the computational domain. Beyond $r > \eta$, the structure functions no longer exhibit power-law behavior. The inset in figure \ref{fig:structure_functions}$b$ shows a semi-logarithmic plot of the second-order structure function (thick curve), compared with $\log(r/\eta)$ over the interval $1\leq r/\eta < 10^2$. In homogeneous and isotropic turbulence, the observed asymptotic scaling $\averagetime{S_2} \sim \log(r/\eta)$, valid for $r\kappa \gg 1$, is consistent with the spectral scaling $\kappa^{-1}$. Moreover, the saturation of the structure function signals the loss of velocity autocorrelation \citep[\S 6]{Pope_2000}. Taken together, these results indicate that heavy Kolmogorov-scale particles locally decorrelate the turbulent flow field over distances exceeding the particle diameter $D$, consistently with the strong attenuation of the energy spectrum for $M_p = 0.6$ (yellow curve in figure \ref{fig:energy_spectra}$a$).

\begin{figure}
  \centering
  \includegraphics[width=\linewidth]{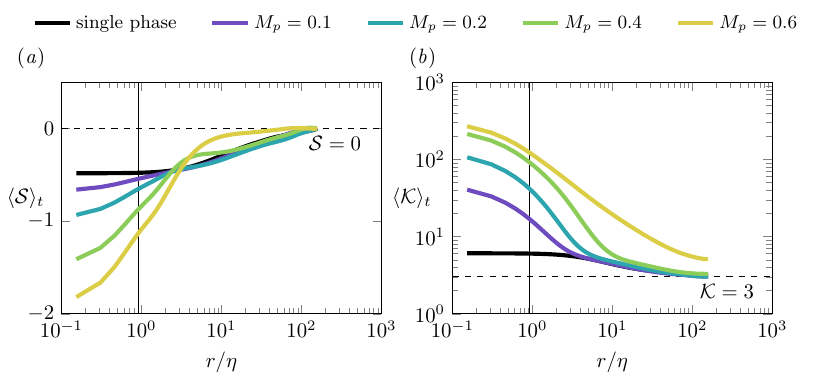}
  \caption{Temporal averages of the skewness $\averagetime{\mathcal{S}}$ ($a$) and kurtosis $\averagetime{\mathcal{K}}$ ($b$) of the velocity field for increasing mass fraction $M_p$. The colours correspond to the cases listed in table \ref{tab:flow_statistics}. The vertical line denotes the particle diameter normalised with the Kolmogorov length $D/\eta$. The horizontal dashed line represent the values $\mathcal{S}=0$ in plot ($a$) and $\mathcal{K}=3$ in plot ($b$).}
  \label{fig:skewness_kurtosis}
\end{figure}

The weakening of the large-scale structures does not result in an attenuation of the turbulent fluctuations nor in the laminarisation of the flow. On the contrary, the intense velocity gradients in the particles' wakes, shown in the visualizations \ref{fig:visualization_dissipation_rate} and \ref{fig:visualization_enstrophy}, indicate enhanced small-scale activity. The deviation of the higher-order moments from the predictions of Kolmogorov theory has been observed in recent PR-DNSs of finite-size and Kolmogorov-size particles \citep{Chiarini_Rosti_2024,Chiarini_Tandurella_Rosti_2025}. To quantify the degree of the flow intermittency (i.e. the relevance of localised events that break the similarity hypothesis), we follow the extended-similarity framework of \cite{Benzi_Ciliberto_Tripiccione_Baudet_Massaioli_Succi_1993} and examine the ratio of the structure functions $S_n/(S_2)^{n/2}$. A departure from the self-similar scaling $S_n/(S_2)^{n/2} = O(1)$ signals the onset of intermittency. We therefore compute the skewness $\mathcal{S} = S_3/S_2^{3/2}$ and the kurtosis $\mathcal{K}=S_4/S_2^{2}$ to investigate the effect of increasing particle inertia on small-scale intermittency. The temporal averages of $\mathcal{S}$ and $\mathcal{K}$ are shown in figure \ref{fig:skewness_kurtosis}$a$ and \ref{fig:skewness_kurtosis}$b$, respectively. For separations $r < D$ (vertical black line in figure \ref{fig:skewness_kurtosis}), the magnitude of $\averagetime{\mathcal{S}}$ and $\averagetime{\mathcal{K}}$ increase monotonically with increasing mass fraction $M_p$, meaning that the deviation from similarity is stronger when heavier particles are considered. This result agrees with the behavior of finite-size particles \citep{Chiarini_Rosti_2024} and Kolmogorov-size particles with $\Phi=10^{-3}$ and $\Psi_p\leq100$ \citep{Chiarini_Tandurella_Rosti_2025}. However, at larger distances, the skewness and kurtosis exhibit opposite behaviour: as $M_p$ increases, the skewness decays more rapidly and approaches the asymptote $\mathcal{S}=0$ (horizontal dashed line), while the kurtosis increases abruptly for $M_p=0.6$ (magenta curves in figure \ref{fig:skewness_kurtosis}) and departs completely from the Gaussian value $\mathcal{K}=3$. In the heaviest-particle case, deviations of $\averagetime{\mathcal{K}}$ are observed at all separations, suggesting that heavy Kolmogorov-scale particles enhance flow intermittency over scales much larger than the particle diameter $D$, thereby altering the inertial cascade and breaking the similarity hypothesis.
\subsection{Scale-by-scale energy budget}\label{sec:scale-by-scale}
The terms of the kinetic energy balance equation provide additional insights into the modulation of the carrier flow. By expressing them as a function of the spatial wavenumber $\kappa$, one can quantify the energy production, transfer and depletion across the scales of motion. The terms are obtained by contracting the Fourier transform of the momentum balance \eqref{eq:NS_momentum} with that of the velocity field $\hat u_i$ \citep{Pope_2000,Cannon_Olivieri_Rosti_2024}. The pressure term vanishes because of \eqref{eq:NS_mass} and the resulting terms are then integrated from a given non-zero wavenumber to infinity, yielding the kinetic energy balance in the spectral domain
\par\vspace{-\baselineskip}%
\begin{multline}\label{eq:kinetic_energy_spectral}
  \frac{\p}{\p t}\int_{\kappa}^\infty \hat E(\kappa) \d\breve\kappa + \underbrace{\frac{1}{2}\int_{\kappa}^\infty \left(\hat G_{i}\hat u_i^\ast + \hat G_{i}^\ast\hat u_i\right) \d\breve\kappa}_{- \Pi(\kappa)} + \underbrace{\int_{\kappa}^\infty \breve\kappa^2 \nu_f \hat u_i\hat u_i^\ast \d\breve\kappa}_{D_v(\kappa)} = \\
  = \underbrace{\frac{1}{2}\int_{\kappa}^\infty \left( \hat f_{i}^{\ABC}\hat u_{i}^\ast + \hat f_{i}^{\ABC\ast}\hat u_i\right) \d\breve\kappa}_{P(\kappa)} + \underbrace{\frac{1}{2}\int_{\kappa}^\infty \left( \hat f_{i}^{\particle}\hat u_i^\ast + \hat f_{i}^{\particle\ast}\hat u_i\right) \d\breve\kappa}_{\Pi_\particle(\kappa)}.
\end{multline}
Here, $\hat G_i = \iunit \kappa_j \widehat{u_ju_i}$ is the Fourier transform of the nonlinear terms, and the superscript $\ast$ denotes the complex conjugate. The integral $\Pi(\kappa)$ is the flux of the nonlinear terms, $\Pi_{\particle}(\kappa)$ the flux of the particle forcing, and $P$ the energy injected by the ABC forcing \eqref{eq:ABC_forcing}. The integral $D_v(\kappa)$ is conveniently expressed as the sum of the bulk dissipation $\averagespace{\varepsilon} = D_v(0)$ and the viscous dissipation integral $\mathcal{D}_v(\kappa)$ \par\vspace{-\baselineskip}%
\begin{equation}
  D_v\left(\kappa\right) = \underbrace{D_v\left(0\right)}_{\averagespace{\varepsilon}} - \underbrace{\int_{0}^\kappa \breve\kappa^2 \nu_f \hat u_i\hat u_i^\ast \d\breve\kappa}_{\mathcal{D}_v(\kappa)}.
\end{equation}
Time-averaging \eqref{eq:kinetic_energy_spectral} yields the scale-by-scale energy budget for a steady-state flow \par\vspace{-\baselineskip}%
\begin{equation}\label{eq:scale-by-scale_energy_budget}
  \averagetime{\Pi\left(\kappa\right) + P\left(\kappa\right) + \Pi_\particle\left(\kappa\right) + \mathcal{D}_v\left(\kappa\right)} = \averagespacetime{\varepsilon}.
\end{equation}
The ABC forcing \eqref{eq:ABC_forcing} is a Dirac delta centred at $\kappa=\kappa_L$ and the integral\par
\begin{figure}
  \centering
  \includegraphics[width=\linewidth]{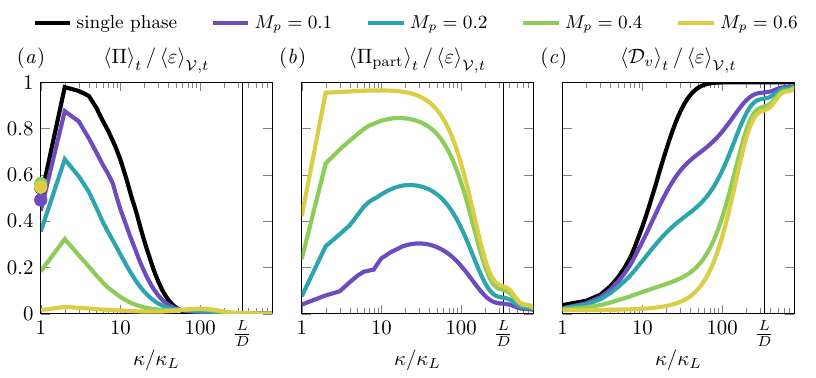}
  \caption{Terms of the normalised scale-by-scale energy budget \eqref{eq:scale-by-scale_energy_budget} for increasing particle inertia. The thick solid curves represent the temporal averages of the turbulent transport term $\averagetime{\Pi}$ (\textit{a}), the particle forcing term $\averagetime{\Pi_{\particle}}$ (\textit{b}) and the dissipation term $\averagetime{\mathcal{D}_v}$ (\textit{c}) normalised on the space-time average of the dissipation rate $\averagespacetime{\varepsilon}$. The ABC forcing $f_i^{\ABC}$ fuels the large-scale motions at $\kappa/\kappa_L=1$ via the integral $\averagetime{P}/\averagespacetime{\varepsilon}$ \eqref{eq:production_integral}, represented by the circular dots \circlemark in panel \textit{a}. The colours correspond to the cases listed in table \ref{tab:flow_statistics}. The black, solid vertical line denotes the wavenumber of the particle diameter $\kappa_D/\kappa_L=L/D$.}
  \label{fig:scalebyscale_energy_transfer}
\end{figure}

\vspace{-\baselineskip}%
\begin{equation}\label{eq:production_integral}
  P\left(\kappa\right) = 
  \begin{aligned}
    &\begin{cases}
      \frac{1}{2}\left(\hat f_{i}^{\ABC}\hat u_i^\ast + \hat f_{i}^{\ABC\ast}\hat u_i\right) & \text{if } \kappa \leq \kappa_L, \\
      0 & \text{if } \kappa > \kappa_L,
    \end{cases}
  \end{aligned}
\end{equation}
has a single non-zero value at $\kappa/\kappa_L=1$ in figure \ref{fig:scalebyscale_energy_transfer}\textit{a} (coloured dots). Hence, for any $\kappa>\kappa_L$, the temporal average of \eqref{eq:scale-by-scale_energy_budget} reduces to $\averagetime{\Pi + \Pi_\particle + \mathcal{D}_v}/\averagespacetime{\varepsilon} = 1$. 

The three panels in figure \ref{fig:scalebyscale_energy_transfer} display the terms of \eqref{eq:scale-by-scale_energy_budget} as functions of $\kappa/\kappa_L$. Here, each curve therefore illustrates the effect of varying $M_p$. The nonlinear transfer $\averagetime{\Pi}$, the work of the ABC forcing $\averagetime{P}$, the work of the particle forcing $\averagetime{\Pi_{\particle}}$, and the viscous dissipation integral $\averagetime{\mathcal{D}_v}$ are all normalised by the mean dissipation $\averagespacetime{\varepsilon}$. The latter is case-dependent (see table \ref{tab:flow_statistics}) and is evaluated at the corresponding mass fraction. As a result, the sum of terms of the same colour across the three panels is equal to unity at all wavenumbers and satisfies the steady-state balance \eqref{eq:scale-by-scale_energy_budget}. For the light-particle cases ($M_p\leq0.2$), the magnitude of the nonlinear flux $\averagetime{\Pi}$ (figure \ref{fig:scalebyscale_energy_transfer}\textit{a}) is essentially the same as in the unladen case, and the effect of the particle forcing $\averagetime{\Pi_\particle}$ (figure \ref{fig:scalebyscale_energy_transfer}\textit{b}) is rather limited. Indeed, the energy transfer is dominated by the nonlinear terms $\averagetime{\Pi}$ and the dissipation $\averagetime{\mathcal{D}_v}$ (figure \ref{fig:scalebyscale_energy_transfer}\textit{c}) at the large scales. This result suggests that the inertial cascade is still active in the range $\kappa/\kappa_L < 20$ for $M_p\leq0.2$, consistently with the $-5/3$ scaling of the spectra in figure \ref{fig:energy_spectra} and the power-law scaling of the second-order structure function in figure \ref{fig:structure_functions}$a$. Figures \ref{fig:scalebyscale_energy_transfer}$a$,$b$ show that the energy flux of the turbulent cascade $\averagetime{\Pi}$ decreases, while the fluid-particle interaction term $\averagetime{\Pi_\particle}$ increases, gradually as particle inertia increases. In the heaviest case ($M_p=0.6$), the contribution of the dissipation and nonlinear terms become negligible at large and intermediate scales, where the energy transfer is dominated by the fluid-particle interaction. The contribution of the nonlinear terms is virtually zero at the particle scale ($\kappa/\kappa_L=L/D$, black vertical line) for all the laden cases. There, $\averagetime{\Pi_\particle}$ is much smaller than $\averagetime{\mathcal{D}_v}$ but increases sensibly with $M_p$. The increase of both $\averagetime{\Pi_\particle}$ and $\averagetime{D_v} = \averagespacetime{\varepsilon} - \averagetime{\mathcal{D}_v}$ at $\kappa\cong\kappa_D$ for increasing $M_p$ was also reported by \citet[see figure 7 therein]{Schneiders_Meinke_Schröder_2017}. These findings are consistent with the relatively small values of the particle Reynolds number $\averageparticletime{\Rey_p}$ (see table \ref{tab:particle_statistics}) and the short, symmetric laminar wakes shown in the insets in figure \ref{fig:visualization_dissipation_rate}$d$ and \ref{fig:visualization_enstrophy}$d$. 

The kinetic energy spectra shown in figure \ref{fig:energy_spectra} and the scale-by-scale energy budget shown in figure \ref{fig:scalebyscale_energy_transfer} support the argument that small, yet heavy particles influence energy transfer at scales much larger than their size. While the presence of Kolmogorov-size particles with moderate $M_p$ mainly influences the local topology of the flow at the smallest scales, high-density suspensions significantly alter the large-scale structures of the flow and enhance the kurtosis of the velocity field across all scales (figure \ref{fig:skewness_kurtosis}$b$). The kinetic energy balance is then dominated by the fluid-particle interaction and the nonlinear fluxes become negligible, significantly departing from the classical Kolmogorov turbulence.

\subsection{Velocity gradient tensor and small-scale intermittency}\label{sec:velocity_gradient_tensor}
The visualizations in figures \ref{fig:visualization_dissipation_rate} and \ref{fig:visualization_enstrophy} depict a pronounced shift in the local distribution of the dissipation rate $\varepsilon$ and the enstrophy $\mathcal{E}$ as particle inertia increases. The peak values of $\varepsilon$ and $\mathcal{E}$ are no longer distributed across large turbulent structures but instead become confined to the vicinity of the particles. Moreover, the modulation of the kinetic-energy spectra and structure functions discussed in \S \ref{sec:energy_spectra_structure_functions}, together with the weakening of the nonlinear energy transfer and the amplification of the fluid--particle interaction reported in \S \ref{sec:scale-by-scale}, suggest an intensification of the small-scale activity. In this section, we examine the properties of the velocity-gradient tensor to gain further insight into the structures of the flow at the particle scale.

The velocity gradient tensor $\p_j u_i$ can be decomposed into its symmetric part, $S_{ij} = (\p_j u_i + \p_i u_j)/2$, and its skew-symmetric part, $W_{ij} = (\p_j u_i - \p_i u_j)/2 = -\levicivita\,\omega_k/2$. The latter is related to the local vorticity $\omega_i = -\levicivita\,\p_j u_k = -\levicivita\,W_{jk}$ \citep{Meneveau_2011,Davidson_2015}. The symmetric part $S_{ij}$ is a diagonal tensor in the frame of reference of its principal axes, the diagonal elements being the three principal rates of strain $\abs{\alpha}\geq\abs{\beta}\geq\abs{\gamma}$, with $\alpha + \beta + \gamma = 0$ for the incompressibility constraint  \eqref{eq:NS_mass}. The eigenvalues of $\partial_j u_i$ satisfy the depressed cubic equation\par\vspace{-\baselineskip}%
\begin{equation}\label{eq:depressed_cubic}
  \lambda^3 + Q\lambda + R = 0,
\end{equation}
where $Q$ and $R$ are the second and the third invariants of $\partial_j u_i$\par\vspace{-\baselineskip}%
\begin{subequations}
\begin{equation}\label{eq:Q_inv}
  Q = - \frac{1}{2}\left(\alpha^2 + \beta^2 + \gamma^2\right) + \frac{1}{4}\left(\omega_x^2+\omega_y^2+\omega_z^2\right),
\end{equation}
\begin{equation}\label{eq:R_inv}
  R = - \alpha\beta\gamma - \frac{1}{4}\left(\alpha\omega_x^2 + \beta\omega_y^2 + \gamma\omega_z^2\right),
\end{equation}
\end{subequations}
respectively. The sign of $Q$ determines whether the local small-scale fluid motion is characterised by strong vorticity ($Q\gg1$) or strong strain ($-Q\gg1$). In the former case, $Q\sim (\omega_x^2 + \omega_y^2 + \omega_z^2)/4$, $R\sim - (\alpha\omega_x^2 + \beta\omega_y^2 + \gamma\omega_z^2)/4$ and the discriminant of \eqref{eq:depressed_cubic} $\Delta = - 4Q^3 - 27R^2 < 0$, meaning that the velocity gradient tensor has one real and two complex conjugate eigenvalues. Positive and negative values of $R$ describe vortex stretching and vortex compression, respectively \citep{Cantwell_1992}. Conversely, when $Q$ is large and negative, the small scales are dominated by axial strain and $R\sim-\alpha\beta\gamma$. Positive and negative values of $R$ describe biaxial strain ($\alpha>0$) and axial strain ($\alpha<0$), respectively. In homogeneous and isotropic turbulence, the average of the invariants $\averagespacetime{Q}$ and $\averagespacetime{R}$ is zero and the global generation of enstrophy can be expressed purely in terms of the axial strain \citep{Betchov_1956}. 

\begin{figure}
  \includegraphics[width=\linewidth]{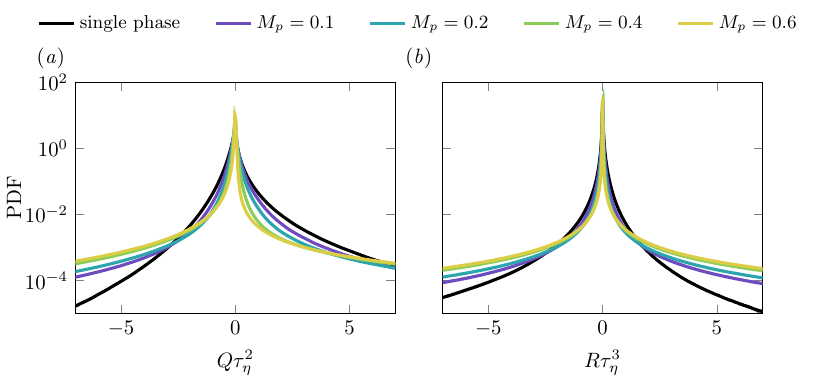}
  \caption{Probability density function (PDF) of ($a$) the second invariant $Q$ \eqref{eq:Q_inv} and ($b$) the third invariant $R$ \eqref{eq:R_inv} of the velocity gradient tensor $\partial_j u_i$ for increasing mass fraction $M_p$. The colours correspond to the cases reported in tables \ref{tab:flow_statistics} and \ref{tab:particle_statistics}. The values of $Q$ and $R$ are normalised with the Kolmogorov time $\tau_\eta$.}
  \label{fig:Q_and_R_PDF}
\end{figure}

Small-scale intermittency drives large negative excursions of $Q$ and positive excursions of $R$ (biaxial strain), leading to sling events that enhance small-particle collisions \citep{Codispoti_Meyer_Jenny_2025}. This behaviour is observed in the probability density functions (PDFs) in figure \ref{fig:Q_and_R_PDF}, where the distributions of the laden cases (coloured curves) are compared to those obtained from an unladen homogeneous and isotropic turbulent flow field. The values of $Q$ and $R$ are scaled with the Kolmogorov time $\tau_\eta$. Increasing particle inertia results in large negative excursions of $Q$ (see figure \ref{fig:Q_and_R_PDF}$a$) and enhances strain at the expense of enstrophy. Note that the positive and negative tails of these distributions also grow when the flow is laden with lighter particles, as previously reported by \cite{Chiarini_Tandurella_Rosti_2025} for $\Psi_p=100$ and $\Phi_p=10^{-3}$. These effects become more and more marked as $\Psi_p$ and $M_p$ increase, and the PDF of $Q$ becomes more negatively skewed as $M_p$ increases (coloured curves \ref{fig:Q_and_R_PDF}$a$). Conversely, the distribution of $R$ becomes progressively symmetric as $M_p$ increases (see figure \ref{fig:Q_and_R_PDF}$b$), thus indicating an equilibrium between axial and biaxial strain and vortex stretching and compression. 

\begin{figure}
  \includegraphics[width=\linewidth]{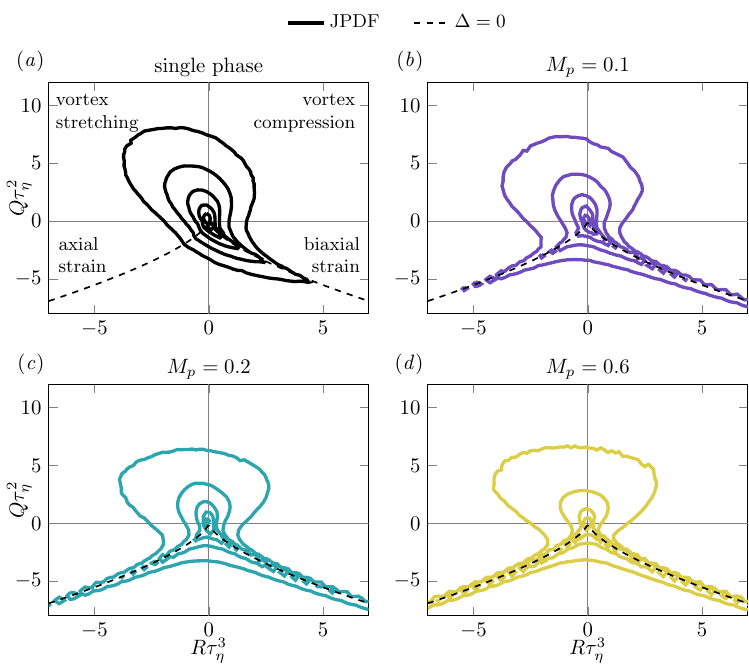}
  \caption{Joint probability density function (JPDF) of the invariants $Q$ and $R$ of the velocity gradient tensor $\partial_j u_i$ for increasing solid mass fraction $M_p$. The invariants are normalised on the Kolmogorov time scale $\tau_\eta$. The five solid curves in each panel represent isolines of contant JPDF with uniform logarithmic spacing (from the outermost $10^{-3}$ to the innermost $10^{1}$). The thin dashed curves describe the two branches of the Vieillefosse tail $Q = - 3(R/2)^{2/3}$, i.e. the points where the discriminant of equation \eqref{eq:depressed_cubic} is zero.}
  \label{fig:QR_JPDF}
\end{figure}

More insights can be obtained from the joint probability density functions (JPDF), shown in figure \ref{fig:QR_JPDF}. The black curves show the contours of constant probability density with uniform logarithmic spacing in the range $10^{{-3}}$ (outermost curve) to $10^{{1}}$ (innermost curve). The fluid phenomena prescribed by the sign of the invariants are summarised in the four quadrants of figure \ref{fig:QR_JPDF}$a$. In the unladen case (figure \ref{fig:QR_JPDF}$a$), vortex stretching is predominant in high-enstrophy regions ($Q>0$, $R<0$), while strain-dominated regions ($Q<0$, $R>0$) are characterised by biaxial strain, as shown by the high-probable events located along the positive-$R$ branch of the Vieillefosse tail, i.e. the zero-discriminant curve $Q = - 3(R/2)^{2/3}$ (blue curve in figure \ref{fig:QR_JPDF}). The shape of the JPDF changes considerably as the density of the suspension increases (figures \ref{fig:QR_JPDF}$b$ and \ref{fig:QR_JPDF}$c$). The distributions become more and more symmetric with respect to the vertical axis $\averagespacetime{R}=0$, although vortex stretching remains slightly predominant in the $Q>0$ region. Strain events become more common as the JPDF spreads along the negative-$R$ branch of the Vieillefosse tail (figures \ref{fig:QR_JPDF}$c$ and \ref{fig:QR_JPDF}$d$), meaning that axial and biaxial strain are equally likely to occur when the flow is laden with heavy particles. 

These results agree with the analysis of \cite{Chiarini_Tandurella_Rosti_2025} but show some discrepancy with those of \cite{Schneiders_Meinke_Schröder_2017}, who investigated a decaying homogeneous and isotropic flow laden with very heavy suspensions (up to $\Psi_p=5000$) at much lower Reynolds numbers ($\Rey_\lambda \leq 50$). In contrast to the present findings, they did not observe the onset of strain-dominated events along the negative-$R$ branch of the Vieillefosse tail, which is associated with an intensification of axial strain and turbulent dissipation. An enhancement of axial strain in the streamline reattachment region is predicted by the Stokes solution for the flow around a sphere or cylinder. It has been reported in PR-DNSs \citep{Chiarini_Tandurella_Rosti_2025} and in wind-tunnel experiments on wavy cylinders at high Reynolds numbers \citep{Ahmed_Khan_Bays-Muchmore_1993}, where it was found to interfere with the pairing of von K\'{a}rm\'an vortices. The enhancement of turbulent dissipation in the particle wakes is also consistent with the amplification of the tails of the PDF of $R$ (figure \ref{fig:Q_and_R_PDF}$b$) and the magnification of the skewness and the kurtosis of the velocity field at small scales (figure \ref{fig:skewness_kurtosis}).


\subsection{Particle velocities and trajectories}\label{sec:particle_motion}

\begin{table}
\centering
\begin{tabular*}{\textwidth}{@{\extracolsep\fill}ccccccccc}
& $\Psi_p$ & $M_p$ & $\displaystyle \frac{\Psi_pD^2}{18\nu_f} \left(\frac{F_0}{L}\right)^{1/2}$ & $\averageparticletime{\Stokes_\eta}$ & $\averageparticletime{\Stokes_\mathcal{L}}$ & $\averageparticletime{\Stokes_{\mathcal{L}0}}$ & $\Rey_p$ & $\displaystyle \frac{\left(\tilde U_i\tilde U_i\right)^{1/2}}{\left(F_0L\right)^{1/2}}$. \\
\midrule
\tikz[baseline=-0.5ex]\draw[fill=mp01, draw=none] (-0.3em,-0.3em) rectangle (1.8em,0.6em); & $100$   & $0.1$ & $0.09$ & $3.84$ & $0.27$ & $0.30$ & $2.18\pm1.23$ & $1.45\pm0.11$ \\
\tikz[baseline=-0.5ex]\draw[fill=mp02, draw=none] (-0.3em,-0.3em) rectangle (1.8em,0.6em); & $250$   & $0.2$ & $0.21$ & $9.14$ & $0.55$ & $0.73$ & $2.81\pm1.49$ & $1.28\pm0.12$ \\
\tikz[baseline=-0.5ex]\draw[fill=mp04, draw=none] (-0.3em,-0.3em) rectangle (1.8em,0.6em); & $666$   & $0.4$ & $0.57$ & $24.10$ & $1.24$ & $1.76$ & $3.83\pm1.81$ & $1.12\pm0.10$ \\
\tikz[baseline=-0.5ex]\draw[fill=mp06, draw=none] (-0.3em,-0.3em) rectangle (1.8em,0.6em); & $1499$  & $0.6$ & $1.31$ & $54.01$ & $2.52$ & $4.03$ & $4.28\pm1.97$ & $1.13\pm0.09$ \\
\end{tabular*}
\caption{Particle parameters and statistics for each case considered in the PR-DNSs. The colours correspond to the cases shown in figures \ref{fig:particle_reynolds_response_time} and \ref{fig:particle_motion}. The volume fraction $\Phi_p = 10^{-3}$ is constant, $\Psi_p=\rho_p/\rho_f$ is the solid-to-fluid density ratio, $M_p$ is the mass fraction defined in \eqref{eq:mass_fraction_def}, $\Psi_pD^2F_0^{1/2}/(18\nu_fL^{1/2})$ the non-dimensional response time for $\Rey_p\ll1$, $\Stokes_\eta = \tau_p/\averagetime{\tau_\eta}$ the Stokes number based on the corrected response time $\tau_p$ \eqref{eq:particle_response_time} and the Kolmogorov time, $\Stokes_\mathcal{L}= \tau_p/\averagetime{\tau_\mathcal{L}}$ the Stokes number based on the eddy turnover time of each laden case and $\Stokes_{\mathcal{L}0} = \tau_p/\averagetime{\tau_{\mathcal{L}0}}$ the Stokes number based on the eddy turnover time of the unladen case. The particle Reynolds number $\Rey_p$ is defined in \eqref{eq:particle_reynolds_number} and $\left(\tilde U_i\tilde U_i\right)^{1/2}\left(F_0L\right)^{-1/2}$ is the non-dimensional particle velocity. Their time-ensemble average $\averageparticletime{\cdot}$ and standard deviation are listed in two the rightmost columns.\label{tab:particle_statistics}}
\end{table}

The modifications of the carrier flow described in the previous sections highlight the effect of small, heavy spheres on the turbulent structures. The latter is manifested through intermittency enhancement (\S \ref{sec:energy_spectra_structure_functions}), energy redistribution (\S \ref{sec:scale-by-scale}) and remodulation of the streamlines at the particle scale (\S \ref{sec:velocity_gradient_tensor}). To further characterise the particle-fluid interphase coupling, we briefly investigate the effect of the mass loading on particle motion in this section. The flow in the vicinity of the particles is characterised by the particle Reynolds number\par\vspace{-\baselineskip}
\begin{equation}\label{eq:particle_reynolds_number}
  \Rey_p = \frac{\left(\Delta\tilde{U}\right) D}{\nu_f},
\end{equation}
based on the velocity difference\par\vspace{-\baselineskip}
\begin{equation}\label{eq:particle_velocity_difference}
  \Delta\tilde U = \left[\left(\tilde U_i - \averageshell{u_i}\right)\left(\tilde U_i - \averageshell{u_i}\right)\right]^{1/2}.
\end{equation} 
Here, the operator $\averageshell{\cdot} = \mathcal{V}_{\shell}^{-1}\int_{\mathcal{V}_\shell}\left(\cdot\right)\d\mathcal{V}$ denotes a volume averaging within a spherical shell centred on the particle and with an inner diameter $D$ and an outer diameter $3D$ \citep{Uhlmann_Chouippe_2017}. Table \ref{tab:particle_statistics} reports the average $\averageparticletime{\cdot}$ and the temporal standard deviation $[\averageparticletime{(\cdot)^2} - \averageparticletime{\cdot}^2]^{1/2}$ of $\Rey_p$. The operator $\averageparticletime{\cdot} = \averagetime{\averageparticle{\cdot}}$ is the ensemble-time average, where $\averageparticle{\cdot} = N^{-1} \sum^N\left(\cdot\right)$ denotes discrete (ensemble) averaging over the $N$ spheres. Indeed, the temporal-ensemble average $\averageparticletime{\Rey_p}$ increases with the mass fraction, its magnitude for the case $M_p=0.6$ being almost twice as much as that for the case $M_p=0.1$. Because $D$ and $\nu_f$ are constant in our simulations, the particle Reynolds number \eqref{eq:particle_reynolds_number} depends only on the velocity difference $\Delta\tilde U_i$. The monotonic increase of $\averageparticletime{\Rey_p}$ with $M_p$ explains the presence of elongated symmetric wakes observed for some of the particles shown in the visualizations \ref{fig:visualization_dissipation_rate} and \ref{fig:visualization_enstrophy}. 

\begin{figure}
  \centering
  \includegraphics[width=\linewidth]{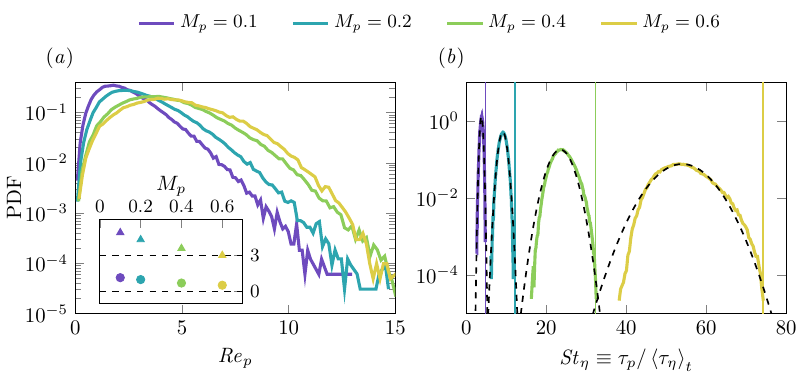}
  \caption{(\textit{a}) PDF of the particle Reynolds number $\Rey_p$ defined in \eqref{eq:particle_reynolds_number}. The colours correspond to the cases reported in table \ref{tab:particle_statistics}. The inset in figure shows the skewness \circlemark and kurtosis \trianglemark of $\Rey_p$ as a function of $M_p$. (\textit{b}) PDF of the Stokes number based on the corrected response time \eqref{eq:particle_response_time} and the Kolmogorov time $\Stokes_\eta\equiv\tau_p/\averagetime{\tau_\eta}$ for increasing $M_p$. The PDFs of $\tau_p/\averagetime{\tau_\eta}$ are compared to the Gaussian distributions of equal mean and standard deviation (black dashed curves), while the vertical coloured lines denote the Stokesian limit of $\Stokes_\eta$ ($\Rey_p\ll1$).}
  \label{fig:particle_reynolds_response_time}
\end{figure}

The PDFs of $\Rey_p$ are shown in figure \ref{fig:particle_reynolds_response_time}\textit{a}. Increasing particle inertia markedly alters the shape of these distributions. For $M_p=0.1$ (violet curve), a large number of particles exhibits relatively small Reynolds numbers, indicating negligible velocity differences with the surrounding fluid $\Delta\tilde U$. This observation is consistent with the visualisations in figures \ref{fig:visualization_dissipation_rate}\textit{b} and \ref{fig:visualization_enstrophy}\textit{b}, which show that particles tend to follow local flow structures. A smaller number of particles decorrelates from the flow and contributes to the pronounced right tail. As $M_p$ increases, the PDFs become less skewed and their peaks shift towards larger $\Rey_p$. The skewness and kurtosis (triangles and circles in the inset) are maximal at $M_p=0.1$ and decrease monotonically with increasing $M_p$. A similar trend is observed for the standard deviation normalised by the mean (see table \ref{tab:particle_statistics}). \cite{Chiarini_Rosti_2024} reported a similar behaviour for suspensions of large spheres ($D/\eta\geq16$) with $\Phi_p=0.08$ and $\Psi_p\leq100$ (refer to figure 16 in their manuscript). While their PDFs spanned much larger values of $\Rey_p=O(10^2)$, their shape is qualitatively similar to the present PDFs. For $M_p=0.6$, the number of particles with $\Rey_p\geq10$ is almost two orders of magnitude higher than at $M_p=0.1$, which is consistent with the elongated wakes in figures \ref{fig:visualization_dissipation_rate}\textit{e} and \ref{fig:visualization_enstrophy}\textit{e}.

The tendency of particles to follow or lag the carrier flow is characterised by the ratio between the particle inertial time scale and that of the carrier flow \citep{Balachandar_Eaton_2010}. In the absence of gravity, particle inertia is described by a response time based on the Stokes drag acting on an isolated sphere \citep{Brandt_Coletti_2022}. However, since $\Rey_p\lesssim10$ in the present simulations, a nonlinear correction such as that of \cite{Schiller_Naumann_1933} is required. Accordingly, we model the integral on the right-hand side of \eqref{eq:newton_euler_momentum} as a nonlinear drag\par\vspace{-\baselineskip}
\begin{equation}
  \oint_{\partial \mathcal{V}_p} \sigma_{ij} n_j \,\mathrm d\left(\partial \mathcal{V}\right) \cong C_{\text{drag}} \rho_f \frac{\left(\Delta\tilde U\right)^2}{2} \pi\left(\frac{D}{2}\right)^2 ,
\end{equation}
with empirical coefficient $C_{\text{drag}} = 24\Rey_p^{-1} (1 + 0.15\Rey_p^{0.687})$ \citep{Hwang_Eaton_2006_nogravity,Balachandar_Eaton_2010,Good_Ireland_Bewley_Bodenschatz_Collins_Warhaft_2014,Ferran_Machicoane_Aliseda_Obligado_2023,van-Wachem_Elmestikawy_Chandran_Hausmann_2025}. Introducing \eqref{eq:particle_reynolds_number}, \eqref{eq:newton_euler_momentum} can be recast as $\mathrm d\tilde U_i/\mathrm dt = \tau_p^{-1}\left(\tilde U_i - \averageshell{u_i}\right)$. Here, the particle response time\par\vspace{-\baselineskip}
\begin{equation}\label{eq:particle_response_time}
  \tau_p = \frac{\Psi_p D^2}{18\nu_f\left(1 + 0.15\Rey_p^{0.687}\right)} 
\end{equation}
reduces to the classical expression $\tau_p = \Psi_p D^2/(18\nu_f)$ for $\Rey_p\ll1$. The time scale $\tau_p$ is normalised by $\averagetime{\tau_\eta}$ and $\averagetime{\tau_{\mathcal{L}}}$, and the corresponding ensemble-time averages are reported in table \ref{tab:particle_statistics}. Both the mean response time \eqref{eq:particle_response_time} and the mean Stokes numbers $\Stokes_\eta \equiv \tau_p/\averagetime{\tau_\eta}$ and $\Stokes_\eta \equiv \tau_p/\averagetime{\tau_\mathcal{L}}$ increase monotonically with $\Psi_p$ and $M_p$. Note that using the eddy turnover time of the unladen case $\averagetime{\tau_{\mathcal{L}0}}$ overestimates the Stokes number. Our values of $\Psi_p$ and $\averageparticletime{\Stokes_\eta}$ are in the same range of those considered in the PR-DNSs of \cite{Schneiders_Meinke_Schröder_2017} and the point-particle DNSs of \cite{Mortimer_Fairweather_2020} and \cite{Gualtieri_Battista_Salvadore_Casciola_2023}, while the averages of $\Rey_p$ on $\Stokes_\eta$ are consistent with the numerical results of \cite{Schneiders_Meinke_Schröder_2017} and \cite{Oka_Goto_2022}. The PDFs of the Stokes number based on the Kolmogorov time scale, $\Stokes_\eta \equiv \tau_p / \averagetime{\tau_\eta}$ (coloured), are compared with Gaussian distributions of the same mean and standard deviation (black dashed) in figure \ref{fig:particle_reynolds_response_time}\textit{b}. The small-$\Rey_p$ limit of $\Stokes_\eta$ is indicated by vertical coloured lines (see also table \ref{tab:particle_statistics}). As particle inertia increases and the distributions of $\Rey_p$ shift towards higher values, the PDFs of $\Stokes_\eta$ flatten and broaden but become narrower than their Gaussian counterparts. Hence, the nonlinear drag is less effective in reducing the response time as $M_p$ increases.

\begin{figure}
  \centering
  \includegraphics[width=\linewidth]{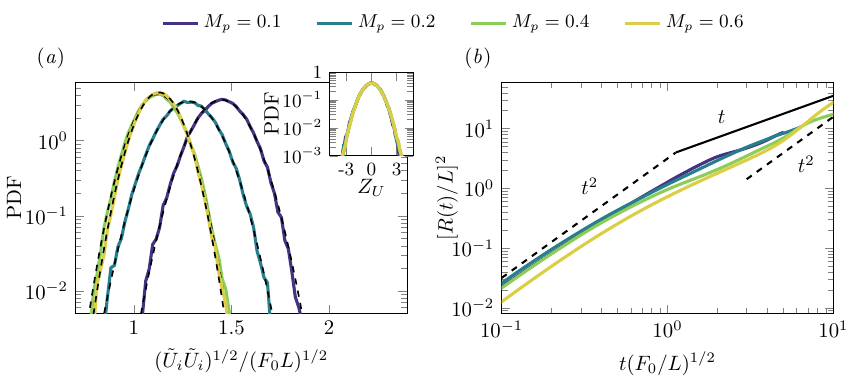}
  \caption{($a$) PDF of the absolute value of the particle velocity $\left(\tilde U_i\tilde U_i\right)^{1/2}$ for increasing particle inertia (thick, coloured curves). The colours correspond to the cases reported in tables \ref{tab:flow_statistics} and \ref{tab:particle_statistics}. The distributions overlap with the Gaussian curves (thin, dashed, black curves). The PDFs are rescaled with respect to the reference velocity $(F_0L)^{1/2}$. The inset shows the PDFs of the normalised variable $Z_{U}$ \eqref{eq:scaled_variable_particle_velocity}. ($b$) Mean square displacement $[R(t)]^2$ of the particles for increasing particle inertia (coloured curves). The time is scaled by a factor $(L/F_0)^{1/2}$ and the displacement by $L$. The results are compared with the scalings $[R(t)]^2 \sim t^2$ (black dashed line) and $[R(t)]^2 \sim t$ (black solid line).}
  \label{fig:particle_motion}
\end{figure}

To confirm that the observed increase in $\Rey_p$ originates from the particle and fluid velocities decorrelation we consider the statistics of the translational velocity of the particles $\tilde U_i$, which is defined in the reference frame of the computational domain. The time and ensemble average of $\left(\tilde U_i\tilde U_i\right)^{1/2}\left(F_0L\right)^{-1/2}$ are listed in the rightmost column in table \ref{tab:particle_statistics}. Figure \ref{fig:particle_motion}$a$ shows the probability density function (PDF) of $\left(\tilde U_i\tilde U_i\right)^{1/2}\left(F_0L\right)^{-1/2}$ computed over the entire time history of each particle for increasing values of the mass fraction $M_p$. For all laden cases (coloured curves), the PDFs closely match Gaussian distributions with the same mean and standard deviation (dashed curves), whose values are reported in the rightmost column of table \ref{tab:particle_statistics}. The normality of the distributions is further confirmed by rescaling the PDFs using the variable \par\vspace{-\baselineskip}%
\begin{equation}\label{eq:scaled_variable_particle_velocity}
  Z_{U} = \left[\left(\tilde U_i\tilde U_i\right)^{1/2} - \averageparticletime{\left(\tilde U_i\tilde U_i\right)^{1/2}}\right] \left[\averageparticletime{\tilde U_i\tilde U_i} - \averageparticletime{\left(\tilde U_i\tilde U_i\right)^{1/2}}^2\right]^{-1/2},
\end{equation}
where we recall that the operator $\averageparticletime{\cdot}$ indicates ensemble and time averaging. The results in the inset of figure \ref{fig:particle_motion}$a$ show an excellent collapse. The mean particle velocity decreases as $M_p$ increases from $0.1$ to $0.4$ but remains approximately constant for $M_p$ between $0.4$ and $0.6$, as indicated by the overlap of the green and yellow curves. Heavier particles are therefore slower in the reference frame of the computational domain. While the velocity difference \eqref{eq:particle_velocity_difference} increases, the particle translational velocity, as well as the fluid kinetic energy and dissipation, remain approximately constant as $M_p$ increases from $0.4$ to $0.6$ (see table \ref{tab:flow_statistics}). This behaviour suggests that the motion of heavy spheres is primarily governed by their inertia and only weakly coupled to the carrier flow. This interpretation is consistent with the reported strain enhancement, with the dissipation and enstrophy peaks near the particles (insets of figures \ref{fig:visualization_dissipation_rate}\textit{e} and \ref{fig:visualization_enstrophy}\textit{e}) and with the enhancement of the fluid velocity skewness and kurtosis at the particle scale (figure \ref{fig:skewness_kurtosis}\textit{b}) when $M_p=0.6$.

For large mass loadings, the present results suggest a weak correlation between particle trajectories and the fluid motion over long times. To characterise such trajectories, we compute the mean-square displacement (i.e.\ the ensemble average of the squared displacement of a particle from its initial position) as a function of time\par\vspace{-\baselineskip}%
\begin{equation}
  [R(t)]^2 = \averageparticle{\abs{X_i(t) - X_i(0)}^2},
\end{equation}
where $X_i(t)$ and $X_i(0)$ denote the particle positions at time $t$ and at the initial time $t=0$, respectively. The results, shown in figure \ref{fig:particle_motion}$b$, reveal a $M_p$-independent ballistic regime $[R(t)]^2 \sim t^2$ at short times $t \ll 1$ (dashed black line), in agreement with previous observations for finite-size spherical particles \citep{Chiarini_Cannon_Rosti_2024}. Here, the time is scaled by a factor $(L/F_0)^{1/2}$ and the displacement by $L$. For times $t = O(1)$, the mean square displacement departs from the ballistic regime. At later times, the cases with $M_p\leq0.4$ approach the Brownian asymptote $\sim t$, although they do not collapse perfectly onto it (solid line). In contrast, the temporal history accessible for the heaviest case ($M_p=0.6$, yellow curve) is not sufficient to identify a clear asymptotic scaling, and the observed trends could therefore reflect a long transient regime. In particular, the heaviest case ($M_p=0.6$) shows a tendency towards $\langle R^2(t)\rangle \sim t^2$ at large times (dashed line), consistent with a ballistic regime in which particle velocities are weakly correlated with the carrier flow. However, establishing whether this behavior is truly asymptotic would require extending the temporal range by at least one additional decade, which is prohibitive for particle-resolved simulations.

\subsection{Particle clustering}\label{sec:particle_clustering}
As shown in the previous section, the motion of the individual spheres decouples from the small-scale turbulent structures for high mass loadings and density ratios. To understand the impact of the inertia of the suspension on the collective motion of the spheres, we proceed by investigating the formation of particle clusters.

Clustering is quantified using a Voronoi tessellation, in which the computational domain is partitioned into $N$ polyhedra -- one associated with each particle -- referred to as Voronoi volumes. The Voronoi volume of a particle consists of all points in space that are closer to that particle than to any other. As a result, smaller Voronoi volumes are found in regions of high particle concentration, whereas larger volumes indicate dilute regions. The Voronoi volume associated with a given particle evolves in time according to its position $X_i$ relative to its neighbours \citep{Monchaux_Bourgoin_Cartellier_2010}, and therefore provides a reliable local measure of the particle concentration \citep{Fiabane_Zimmermann_Volk_Pinton_Bourgoin_2012}. Owing to its locality in both space and time, the Voronoi tessellation captures the spatial structure of particle clusters, as well as their temporal evolution \citep{Uhlmann_Chouippe_2017}. In the present study, we computed the three-dimensional Voronoi tessellation using an in-house code based on the C++ library \href{https://math.lbl.gov/voro++/}{Voro++} \citep{Lu_Lazar_Rycroft_2023}.
\begin{figure}
  \centering
  \includegraphics[width=\linewidth]{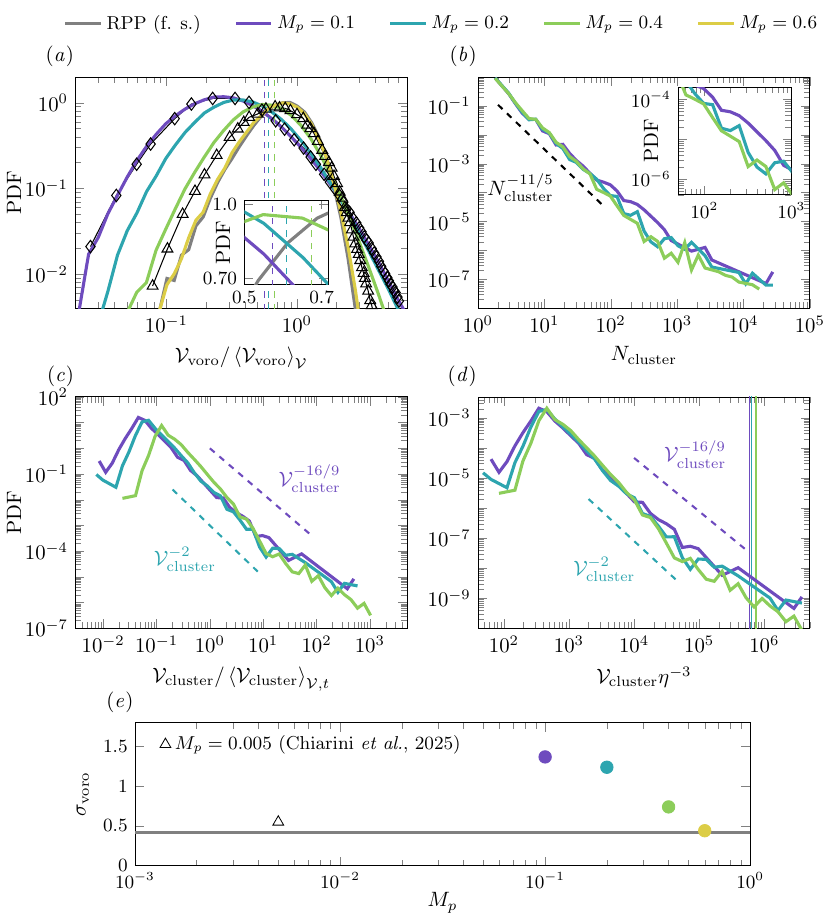}
  \caption{($a$) PDFs of the normalised Voronoi volumes $\vvoro/\averagespace{\vvoro}$ for increasing particle-to-fluid density ratio $\Psi_p$ and mass fraction $M_p$ (coloured curves) are compared to that obtained for a suspension of finite-size (f. s.) particles distributed via a random-Poisson process (RPP, gray). The PDFs computed by \cite{Chiarini_Tandurella_Rosti_2025} are drawn with the triangles \trianglemark ($\Psi_p=5$, $M_p=0.005$) and the diamonds \diamondmark ($\Psi_p=100$, $M_p=0.1$). The inset shows the position of the lower cross-over points (LCO) listed in table \ref{tab:voronoi_statistics} (dashed vertical lines). ($b$) PDF of the number of particles forming a cluster $\Ncluster$. The inset shows a zoomed view of the PDFs in the range $50<\Ncluster<10^3$. ($c$) PDF of the cluster volumes normalised on the space-time average $\averagespacetime{\Vcluster}$ (solid lines), along with the scaling $-2$ and $-16/9$ (dashed lines). ($d$) PDF of the cluster volumes normalised on $\eta^3$ (solid lines), along with the scaling $-2$ and $-16/9$ (dashed lines). The solid vertical lines denote a normalised cluster volume equal to the cube of the integral scale $\averagetime{\mathcal{L}}^3\averagetime{\eta}^{-3}$. (\textit{e}) Standard deviation of the Voronoi volumes $\sigmavoro$ for increasing $M_p$. The data of the present simulations (\circlemark) are compared with the case $M_p=0.005$ computed by \citet[figure 18\textit{b} therein \trianglemark]{Chiarini_Tandurella_Rosti_2025} and with the RPP baseline (gray horizontal line). The colours correspond to the cases listed in table \ref{tab:voronoi_statistics}.}
  \label{fig:voronoi}
\end{figure}

The tendency of particles to form clusters is examined by comparing the probability density functions (PDFs) of the Voronoi volumes obtained from the PR-DNSs with those computed for a population of finite-size particles whose spatial coordinates are assigned through a random Poisson process (RPP). The time-averaged PDFs of the Voronoi volumes $\vvoro$ obtained from the DNSs are plotted for increasing mass fraction $M_p$ (coloured curves) in figure \ref{fig:voronoi}$a$. The PDF of a population of $N$ non-overlapping, Kolmogorov-size particles distributed through a RPP is plotted in black for comparison. All the PDFs in figure \ref{fig:voronoi}$a$ were rescaled on the inverse of the global concentration $\averagespace{\vvoro}=L^3/N=3.35\times10^{-3}$ which is time-independent \citep{Fiabane_Zimmermann_Volk_Pinton_Bourgoin_2012}. The departure of the PDFs from the RPP case is maximum in the light-particle case $M_p=0.1$ (violet curve), while the PDFs approach the RPP case gradually as $\Psi_p$ and $M_p$ increase, ultimately collapsing on the RPP case for $M_p=0.6$. These PDFs have unit mean and their broadening indicates the presence of Voronoi volumes that are either smaller (concentrated regions) and larger (dilute regions) than those computed for the RPP. \cite{Monchaux_Bourgoin_Cartellier_2010} showed that the PDFs of Voronoi areas of two-dimensional tessellations are well-approximated by log-normal distributions, whereas \cite{Tagawa_Martinez-Mercado_Prakash_Calzavarini_Sun_Lohse_2012} pointed out that a Gamma distribution better interpolates the PDFs of Voronoi volumes in three-dimensional tessellations. Either way, the log-normal and the Gamma distributions that fit the PDFs of $\vvoro/\averagespace{\vvoro}$ are characterised by the standard deviation $\sigmavoro$, which quantifies the degree of clustering \citep{Monchaux_Bourgoin_Cartellier_2010}. 

\begin{table}
\centering
\begin{tabular*}{\textwidth}{@{\extracolsep\fill}cccccccccc}
& $\Psi_p$ & $M_p$ & $\sigmavoro$ & $\skewvoro$ & $\kurtvoro$ & LCO & HCO & $\averagespacetime{\Vcluster}^{1/3}\averagetime{\eta}^{-1}$ & $\averagespacetime{\Ncluster}$ \\
\midrule
\tikz[baseline=-0.5ex]\draw[fill=mp01, draw=none] (-0.3em,-0.3em) rectangle (1.8em,0.6em); & $100$   & $0.1$ & $1.37$ & $5.58$ & $56.20$ & $0.57$ & $1.96$ & $19.62$ & $54$ \\
\tikz[baseline=-0.5ex]\draw[fill=mp02, draw=none] (-0.3em,-0.3em) rectangle (1.8em,0.6em); & $250$   & $0.20$ & $1.24$ & $6.18$ & $71.15$ & $0.60$ & $1.95$ & $18.60$ & $41$ \\
\tikz[baseline=-0.5ex]\draw[fill=mp04, draw=none] (-0.3em,-0.3em) rectangle (1.8em,0.6em); & $666$   & $0.40$ & $0.74$ & $3.19$ & $24.87$ & $0.67$ & $1.82$ & $15.42$ & $19$ \\
\tikz[baseline=-0.5ex]\draw[fill=mp06, draw=none] (-0.3em,-0.3em) rectangle (1.8em,0.6em); & $1499$  & $0.60$ & $0.44$ & $0.93$ & $4.37$ & n.a. & n.a. & n.a. & n.a. \\
\tikz[baseline=-0.5ex]\draw[fill=gray, draw=none] (-0.3em,-0.3em) rectangle (1.8em,0.6em); & \multicolumn{2}{c}{RPP (finite size)} & $0.42$ & $0.80$ & $3.90$ & n.a. & n.a. & n.a. & n.a. \\
\end{tabular*}
\caption{Properties of the Voronoi tessellation. The colours correspond to the cases shown in figures \ref{fig:voronoi} and \ref{fig:preferential_sampling_PDF_Q}.  The table lists the standard deviation $\sigma$, skewness $\mathcal{S}$, and kurtosis $\mathcal{K}$ of the normalised Voronoi volumes $\vvoro/\averagespace{\vvoro}$, along with the lower cross-over points (LCO) and the higher cross-over points (HCO) of their PDFs. The two rightmost columns list the characteristic cluster length $\averagespacetime{\Vcluster}^{1/3}\averagetime{\eta}^{-1}$ and the average number of particles forming a cluster $\averagespacetime{\Ncluster}$. All variables are non-dimensional. \label{tab:voronoi_statistics}}
\end{table}

The statistical properties of the Voronoi tessellation are listed in table \ref{tab:voronoi_statistics}. The standard deviation $\sigmavoro$ is maximum for $M_p=0.1$ ($\averageparticletime{\Stokes_\eta}=3.84$, violet curve) and decreases monotonically as the PDFs of the Voronoi volumes approach the RPP case (gray curve in figure \ref{fig:voronoi}$a$) for large $M_p$. The probability of large-concentration events is therefore maximum for $M_p=0.1$ and decreases for larger $M_p$. While the PDFs for $M_p=0.2$ and $M_p=0.4$ reveal a moderate presence of clustering events, the distribution for the heaviest-particle case $M_p=0.6$ ($\averageparticletime{\Stokes_\eta}=54.01$, yellow curve) collapses on that of the RPP, meaning that clustering does not occur in this case. Similarly, the skewness $\skewvoro$ and the kurtosis $\kurtvoro$ are maximum for $M_p=0.2$ and approach the RPP values for $M_p=0.6$. Although the degree of clustering (i.e. $\sigmavoro$) decreases for decreasing $\Rey_\lambda$ and increasing $\Stokes_\eta$, separating the effect of each parameter is not straightforward \citep{Hassaini_Petersen_Coletti_2023}. \cite{Sumbekova_Cartellier_Aliseda_Bourgoin_2017} measured the clustering of polydisperse smaller-than-Kolmogorov particles in homogeneous and isotropic turbulence and reported a weak influence of $\Stokes_\eta$ and a strong influence of $\Rey_\lambda$ and $\Phi_p$ on $\sigmavoro$. However, the DNS data reviewed by \cite{Monchaux_2012} reveal a marked influence of the Stokes number on the degree of clustering. In the present work, increasing $M_p$ from 0.4 and 0.6 results in the reduction of $\sigmavoro$ from $0.74$ to $0.44$ and in the increase of $\averageparticletime{\Stokes_\eta}$ from $24.10$ to $54.01$, but does not affect $\Rey_\lambda$ sensibly (see tables \ref{tab:voronoi_statistics}, \ref{tab:particle_statistics} and \ref{tab:flow_statistics}). 

For an unladen Taylor-Reynolds number $\Rey_\lambda\cong140$ and the same volume fraction $\Phi_p=10^{-3}$, \cite{Chiarini_Tandurella_Rosti_2025} computed the Voronoi tessellation and reported enhanced clustering as the density ratio $\Psi_p$ increased from 5 to 100. Their PDFs are reproduced in figure \ref{fig:voronoi}\textit{a} for $M_p=0.005$ ($\Psi_p=5$, triangles) and $M_p=0.1$ ($\Psi_p=100$, diamonds), the latter overlapping with the present results (violet curve). To confirm this trend, we estimate $\sigmavoro$ for $M_p=0.005$ from its data and plot it alongside the values obtained in the present simulations in figure \ref{fig:voronoi}\textit{e}. The degree of clustering increases for $M_p<0.1$ and decreases for $M_p>0.1$. When clustering peaks, the average Stokes number based on the Kolmogorov time scale is slightly larger than unity $\averageparticletime{\Stokes_\eta}=3.84$ (refer to figure \ref{fig:particle_reynolds_response_time}$b$). While the PDFs of \cite{Chiarini_Tandurella_Rosti_2025} shift towards smaller Voronoi volumes as $\Psi_p$ increases from 5 to 100, the present results indicate the opposite trend when $\Psi_p$ is further increased to values typical of gas-solid systems, $O(10^3)$ \citep{Rose_Durant_2009}, suggesting a weakening of clustering in this regime.

We follow the criterion proposed by \cite{Monchaux_Bourgoin_Cartellier_2010} to study the spatial extension of the particle clusters. The PDFs obtained from the PR-DNSs intersecate that of the RPP case for two values of $\vvoro/\averagespace{\vvoro}$. Voronoi cells smaller than the low cross-over points (LCO) are flagged as dense and may be part of a cluster. Conversely, cells greater than the high cross-over point (HCO) belong in void regions. The LCO points and the HCO points are listed in table \ref{tab:voronoi_statistics} and are denoted by the dashed vertical lines in figure \ref{fig:voronoi}$a$ (see also the inset for an enlarged view). Although two cross-over points exist in the case $M_p=0.6$, the clustering criterion of \cite{Monchaux_Bourgoin_Cartellier_2010} can not be applied because the curves of the RPP case and the heaviest case ($M_p=0.6$) almost overlap and their normalised standard deviation is approximately the same (see table \ref{tab:voronoi_statistics}). Thus, we discuss clustering for the cases $M_p=0.1$, $0.2$ and $0.4$. 

The condition $\vvoro/\averagespace{\vvoro} < \text{LCO}$ is necessary but not sufficient for a particle to be considered part of a cluster \citep{Baker_Frankel_Mani_Coletti_2017}. Accordingly, only dense Voronoi cells that share at least one vertex are regarded as part of the same cluster \citep{Hassaini_Petersen_Coletti_2023}. For each dense cell, its neighbours are examined to identify adjacent dense cells, which are then grouped into a single cluster. A minimum of two contiguous dense cells is required to define a cluster, and a cell cannot belong to multiple clusters. To prevent the unphysical merging of adjacent yet distinct clusters, dense-cell particles separated by a distance larger than the cut-off $\averagespace{\vvoro}^{1/3}$ are excluded.

The PDFs of the number of particles forming a cluster $\Ncluster$ are shown in figure \ref{fig:voronoi}$b$. The distributions are in qualitative agreement with those of \cite{Uhlmann_Chouippe_2017} (refer to figure 10$a$ therein) who performed PR-DNS of spherical particles with $D/\eta=5$ in homogeneous and isotropic turbulence. These PDFs collapse for $\Ncluster<10^2$ where they show a power-law dependency $\Ncluster^{-11/5}$ which is not influenced by particle inertia. They depart for larger $\Ncluster$, meaning that large clusters are less likely to form as $M_p$ increases (see inset in figure \ref{fig:voronoi}$b$). The average number of particles per cluster, listed in table \ref{tab:voronoi_statistics}, diminishes markedly as $M_p$ increases.

The PDFs of the cluster volumes $\Vcluster$ are shown in figure \ref{fig:voronoi}$c$ and \ref{fig:voronoi}$d$, where they are rescaled on the average $\averagespacetime{\Vcluster}$ \citep{Sumbekova_Cartellier_Aliseda_Bourgoin_2017} and the cube of the Kolmogorov scale $\eta$ \citep{Uhlmann_Chouippe_2017,Hassaini_Petersen_Coletti_2023}. These PDFs exhibit a distinct peak followed by a power-law decay which suggests the existence of a characteristic cluster size. The characteristic length $\averagespacetime{\Vcluster}^{1/3}$ \citep{Rostami_Li_Kheirkhah_2024} ranges between $10\eta$ and $20\eta$ (see table \ref{tab:voronoi_statistics}), in close agreement with laboratory measurements of polydisperse distributions \citep{Obligado_Teitelbaum_Cartellier_Mininni_Bourgoin_2014,Sumbekova_Cartellier_Aliseda_Bourgoin_2017}, monodisperse distributions \citep{Hassaini_Petersen_Coletti_2023}, and with the numerical results of \cite{Uhlmann_Chouippe_2017}. The PDFs in figure \ref{fig:voronoi}$c$ and \ref{fig:voronoi}$d$ exhibit a power-law behaviour spanning approximately two decades, suggesting a self-similarity in the cluster structure consistent with the theory of \cite{Goto_Vassilicos_2006}. The power-law region sets in for volumes larger than $\averagespacetime{\Vcluster}$, and features a slope close to $-16/9$ for the case $M_p=0.1$ (blue) and to $-2$ for $M_p=0.2$ (orange). For $M_p=0.1$, the power-law region extends up to volumes $O(\mathcal{L}^3)$, $\mathcal{L}$ being the integral scale (vertical solid lines in figure \ref{fig:voronoi}$d$). The exponent $-16/9$ was predicted by \cite{Yoshimoto_Goto_2007}, who extended the fractal model of \cite{Goto_Vassilicos_2006} to the three-dimensional case. This model describes the statistical distribution of cluster areas in two-dimensional isotropic turbulence, and relies on the assumption of a fully developed inertial subrange, as well as on the correspondence between void areas and self-similar eddies at discrete scales. In the two-dimensional case, the model predicts the PDF for the cluster areas to decay with a slope $-5/3$. \cite{Sumbekova_Cartellier_Aliseda_Bourgoin_2017} performed wind-tunnel measurements on sub-Kolmogorov particles at $\Rey_\lambda \leq 400$ and reported a $-5/3$ scaling after normalising each PDF by its temporal average. \cite{Uhlmann_Chouippe_2017}, reported a slope $-16/9$, in good agreement with the theory of \cite{Yoshimoto_Goto_2007}. \cite{Baker_Frankel_Mani_Coletti_2017} performed point-particle DNS of smaller-than-Kolmogorov spherical particles in homogeneous isotropic turbulence, and reported a slope close to $-2$ for large $\Vcluster$. More recently, \cite{Hassaini_Petersen_Coletti_2023} computed the PDFs of the cluster areas from particle-image-velocimetry measurements of a turbulent flow field laden with smaller-than-Kolmogorov particles. 

The PDFs in figure \ref{fig:voronoi}$c$ and \ref{fig:voronoi}$d$ do not exhibit a clear broadening of the self-similar clustering range when $M_p$ and $\Stokes_\eta$ increase, $\Rey_\lambda$ decreases, and $\Phi_p$ is held constant. This behaviour contrasts with the experimental findings of \citet[see figure 3 therein]{Hassaini_Petersen_Coletti_2023}, who reported a widening of the self-similar range with increasing $M_p$ and $\Phi_p$, while keeping $\Psi_p$, $\Stokes_\eta$, and $\Rey_\lambda$ constant for monodisperse sub-Kolmogorov particles, thus suggesting a complex interaction among the various parameters governing the flow. The PDFs of $\Vcluster\eta^{-3}$ (figure \ref{fig:voronoi}$d$) show that only the very small clusters ($\Vcluster^{1/3}<8\eta$) and the large clusters of length $\Vcluster^{1/3}>\average{\Vcluster}^{1/3}$ are affected by increasing particle inertia. Comparison of figures \ref{fig:voronoi}$b$,$d$ shows that, as $\Psi_p$ and $M_p$ increase, the number of particles in the smallest clusters remains constant, but their Voronoi volumes shrink, while larger clusters are populated by fewer particles. 

In summary, clustering attenuation is accompanied by a reduction of $\Rey_\lambda$ and by the departure of the energy spectra from a well-defined self-similar scaling. When clustering is maximal, the Stokes number based on the Kolmogorov time scale exhibits a relatively narrow distribution and is slightly larger than unity, suggesting that the spheres resonate with fluid structures slightly larger than the Kolmogorov scale. Although the mean cluster length lies near the high-wavenumber limit of the inertial subrange ($\kappa/\kappa_L=16$), the particles also form larger clusters whose sizes follow a power-law distribution (see figure \ref{fig:voronoi}$c,d$). This behavior changes markedly as the particle inertia increases: the nonlinear energy transfer weakens, the flow structures decorrelate over separations larger than $\eta$ and clustering disappears. The heaviest particles neither cluster nor interact significantly with larger turbulent structures and are characterised by much larger values of $\Stokes_\eta = O(10^2)$.

\subsection{Preferential concentration}\label{sec:preferential_sampling}

\begin{figure}
  \centering
  \includegraphics[width=\linewidth]{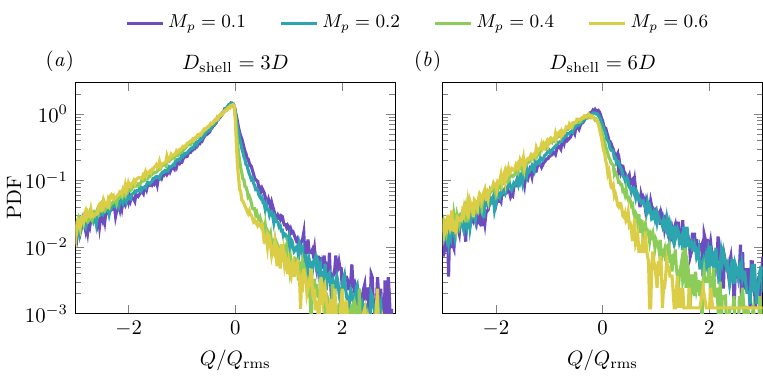}
  \caption{Probability density function (PDF) of the second invariant of the velocity gradient tensor normalised on its root-mean-square value $Q/Q_{\text{rms}}$, evaluated at the particle position for increasing $M_p$. The colours correspond to the cases reported in table \ref{tab:voronoi_statistics}. The normalised invariant $Q$ is evaluated within a spherical shell with an external diameter ($a$) $D_\shell=3D$ and ($b$) $D_\shell=6D$.}
  \label{fig:preferential_sampling_PDF_Q}
\end{figure}

The results of \S\ref{sec:particle_clustering} suggest a connection between particle clustering and the presence of a well-developed inertial subrange and point to the role of turbulent structures in the formation of clusters. In this section, we investigate whether the motion of heavy Kolmogorov-size particles is influenced by the centrifugal mechanism proposed by \cite{Maxey_1987}. Following \cite{Squires_Eaton_1990} and \cite{Chiarini_Tandurella_Rosti_2025}, we compute the probability density function (PDF) of the second invariant of $\partial_j u_i$ \eqref{eq:Q_inv}, normalised by its root-mean-square value, $Q/Q_{\text{rms}}$, and evaluated in the vicinity of the particle. This local value of $Q$ is estimated by volume averaging over a spherical shell with inner diameter $D$ and outer diameter $D_{\shell}$. The resulting statistics depend on the choice of $D_{\shell}$: if the shell is too thin, the averaging primarily captures shear stresses at the particle surface; if it is too thick, spurious contributions from regions far from the particle are included \citep{Uhlmann_Chouippe_2017,Oka_Goto_2022}. The effect of varying $D_{\shell}$ was previously investigated by \cite{Chiarini_Tandurella_Rosti_2025} (see figure 20 therein), who showed that Kolmogorov-size spheres of moderate inertia ($\Psi_p=5$, $100$) preferentially sample regions of negative $Q$ (i.e. where strain overcomes vorticity). In the present work, we compute the averaged $Q$ using $D_{\shell} = 3D$ \citep{Uhlmann_Chouippe_2017} and compare the results with those obtained using $D_{\shell} = 6D$. The effect of particle inertia on the PDFs of $Q/Q_{\text{rms}}$ is shown in figure \ref{fig:preferential_sampling_PDF_Q}$a$ ($D_\shell=3D$) and \ref{fig:preferential_sampling_PDF_Q}$b$ ($D_\shell=6D$) for increasing $M_p$. 

Regardless of the value of $D_\shell$, the PDFs are negatively skewed in all cases. The degree of clustering (see table \ref{tab:voronoi_statistics}) is maximal for $M_p=0.1$ and decreases monotonically with increasing mass loading, suggesting that preferential sampling is most relevant when $\Stokes_\eta = O(1)$. The left tail increases and the right tail decreases as $M_p$ increases from $0.1$ to $0.4$. Although this trend suggests that Kolmogorov-scale particles preferentially sample strain-dominated regions and avoid vortical ones, a similar behavior is also observed in the PDFs of $Q$ computed over the whole domain (figure \ref{fig:Q_and_R_PDF}$a$). Moreover, unlike point-particle DNSs, PR-DNSs do not allow for a clear separation between preferential sampling and flow modulation. It is therefore difficult to determine whether this trend originates from modifications of the carrier flow or from enhanced preferential sampling as $M_p$ increases.
 
The case $M_p=0.6$ (magenta curves in figure \ref{fig:preferential_sampling_PDF_Q}) does not feature a canonical Kolmogorov cascade and thus deserves more attention. The regions of very high dissipation rate and enstrophy shown in the insets in the visualizations \ref{fig:visualization_dissipation_rate}$d$ and \ref{fig:visualization_enstrophy}$d$, respectively, are located in the shear layers and in the symmetric laminar wakes that appear when the average particle Reynolds number is $\averageparticletime{\Rey_p}\gtrsim1$ (see table \ref{tab:particle_statistics}). A possible explanation is that, as particle inertia increases, the nonlinear transfer of kinetic energy $\Pi$ vanishes and the energy balance \eqref{eq:scale-by-scale_energy_budget} is mediated by the fluid-particle interaction term $\Pi_\particle$ and the dissipation $D_v$. The intermediate- and small-scale eddies weaken, the peaks of $\varepsilon$ and $\mathcal{E}$ become localised in the particles' shear layers, and clustering is no longer observed. Hence, heavy Kolmogorov-size particles sample regions of high strain and low enstrophy, but the centrifugal mechanism may affect their trajectory only in the presence of a fully-developed inertial subrange and for moderate values of $\Psi_p$. The predominance of negative values of $Q/Q_{\text{rms}}$ for $M_p=0.6$ should therefore be attributed to the local increase of axial strain in the particle wakes (see figure \ref{fig:QR_JPDF}$d$), rather than interpreted as evidence of enhanced preferential sampling.

\begin{figure}
  \centering
  \includegraphics[width=\linewidth]{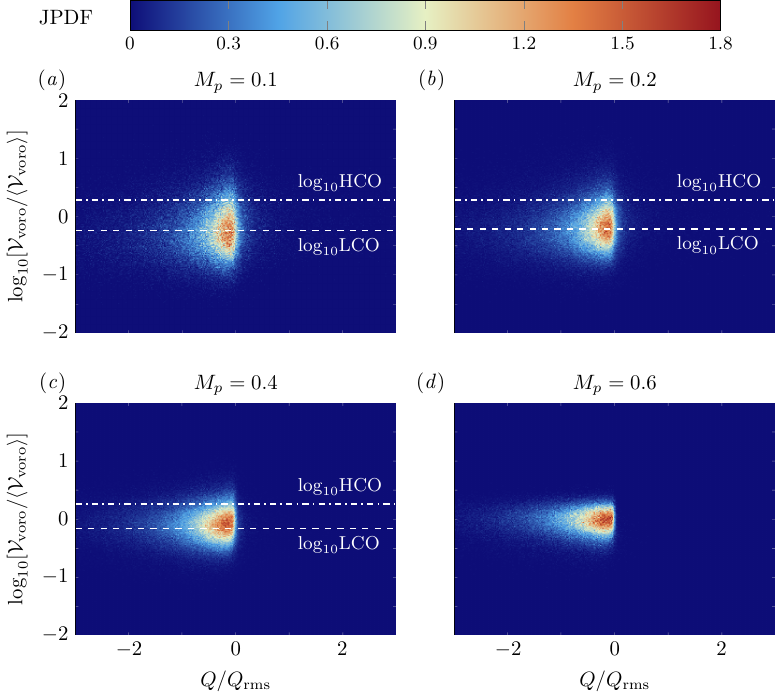}
  \caption{Joint probability density function (JPDF) of the normalised second invariant of the velocity gradient tensor $Q$ \eqref{eq:Q_inv} evaluated around the particles, and the normalised Voronoi volume $\vvoro/\averagespace{\vvoro}$ for ($a$) $M_p=0.1$, ($b$) $M_p=0.2$, ($c$) $M_p=0.4$ and ($d$) $M_p=0.6$. The magnitude of the JPDF is described by the bar at the top of the four panels. The position of the high cross-over (HCO) and low cross-over (LCO) points are plotted with the white dash-dotted and white dashed lines, respectively (refer also to table \ref{tab:voronoi_statistics}). The second invariant $Q$ was computed by averaging over a spherical shell of diameter $D_{\text{shell}}=3D$ around the particle.}
  \label{fig:preferential_sampling_JPDF_Q_voronoi}
\end{figure}

To gain further insights on the interplay between particle clustering and preferential concentration, we compute the joint probability density function (JPDF) of the normalised Voronoi volumes $\vvoro/\averagespace{\vvoro}$ and the normalised second invariant of the velocity tensor evaluated at the particle position $Q/Q_{\text{rms}}$ \citep{Tagawa_Martinez-Mercado_Prakash_Calzavarini_Sun_Lohse_2012}. The results are shown in figure \ref{fig:preferential_sampling_JPDF_Q_voronoi} for ($a$) $M_p=0.1$, ($b$) $M_p=0.2$, ($c$) $M_p=0.4$ and ($d$) $M_p=0.6$. The values of $\vvoro/\averagespace{\vvoro}$ corresponding to the low and high cross-over points (LCO and HCO) are also shown as white dashed and dash-dotted lines, respectively. Once again, we computed these joint distributions by volume-averaging $Q$ within a spherical shell of outer diameter $D_\shell=3D$ around the particles \citep{Uhlmann_Chouippe_2017,Chiarini_Tandurella_Rosti_2025}. Voronoi volumes smaller than the $\text{LCO}$ threshold correlate with negative values of $Q/Q_{\text{rms}}$ for the cases $M_p=0.1$ and $M_p=0.2$. In these cases, the LCO threshold intersects the peak of the joint distribution (red region in figures \ref{fig:preferential_sampling_JPDF_Q_voronoi}$a$,$b$). Thus, the particles that are most likely to be part of a cluster lie in high-strain, low-enstrophy regions of negative $Q$. A similar conclusion can be drawn for the case $M_p=0.4$, although the peak of the joint distribution now lies between the HCO and LCO thresholds and the amount of the particles with $\vvoro/\averagespace{\vvoro}<\text{LCO}$ is smaller. This result is consistent with the steep decrease of the cluster characteristic length $\averagespacetime{\Vcluster}^{1/3}$ and the number of particles per cluster $\averagespacetime{\Ncluster}$ that we observe when the mass fraction increases from $0.2$ to $0.4$ (see table \ref{tab:voronoi_statistics}). While \cite{Chiarini_Tandurella_Rosti_2025} showed that the joint probability of negative $Q$ and small Voronoi volumes increases as $\Psi_p$ increased from $5$ to $100$, figure \ref{fig:preferential_sampling_JPDF_Q_voronoi} shows that the opposite occurs when particle inertia increases beyond $\Psi_p=100$. The present results are in qualitative agreement with those of \citet[see figure 7 therein]{Tagawa_Martinez-Mercado_Prakash_Calzavarini_Sun_Lohse_2012}, who performed point-particle DNSs and reported the joint PDF of the normalised Voronoi volumes and enstrophy at $\Rey_\lambda=75$, $\Stokes_\eta=0.6$, and in the limits $\Psi_p\ll1$ and $\Psi_p\gg1$. They observed a stronger correlation between small Voronoi volumes and large enstrophy for small $\Psi_p$ — consistent with PR-DNS results \citep{Chiarini_Tandurella_Rosti_2025} — as well as enhanced clustering in the large-$\Psi_p$ limit. The discrepancies at large $\Psi_p$ are likely attributable to the lower $\Rey_\lambda$ and to the inherent limitations of the point-particle approximation.

\section{Conclusions}\label{sec:conclusions}
We investigate homogeneous and isotropic turbulence laden with a dilute suspension of heavy, Kolmogorov-scale particles. To quantify the particle-fluid feedback, we perform particle-resolved direct numerical simulations (PR-DNSs) in a triperiodic cubic domain containing a monodisperse population of $74\,208$ solid spheres. Fluid-solid interactions are resolved using the Eulerian immersed boundary method (IBM) of \cite{Hori_Takagi_Rosti_2022}. The particle diameter, domain size, and fluid properties are held fixed to isolate the effects of increasing particle-to-fluid density ratio and mass fraction at constant volume fraction $\Phi_p = 10^{-3}$. Four density ratios are considered, $\Psi_p = 100$, $250$, $666$, and $1499$, corresponding to mass fractions $M_p = 0.1$, $0.2$, $0.4$, and $0.6$, respectively. In the unladen reference case, a Taylor-scale Reynolds number $\Rey_\lambda \cong 150$ is achieved, ensuring adequate scale separation and the presence of an inertial subrange extending over around one decade of wavenumbers.

The bulk properties of the carrier flow are substantially modified once the particle-to-fluid density ratio exceeds $\Psi_p = 100$. As particle inertia increases, the dispersed phase extracts a significant amount of the kinetic energy from the carrier flow, leading to a marked reduction of the Taylor-scale Reynolds number. The bulk dissipation rate is also affected. However, the Kolmogorov length scale remains approximately constant and comparable to the particle diameter across all cases. The energy spectra exhibit a progressive departure from the classical Kolmogorov $\kappa^{-5/3}$ scaling as the mass fraction increases. When the flow is laden, the spectra display a $\kappa^{-4}$ decay at high wavenumbers, corresponding to sub-Kolmogorov and sub-particle scales. This scaling is reminiscent of those reported in bubbly flows and has recently been observed in particle-resolved simulations by \cite{Chiarini_Tandurella_Rosti_2025}. At low and intermediate wavenumbers, the classical Kolmogorov scaling is replaced by a $\kappa^{-1}$ regime when the flow is laden with the heaviest particles. This behaviour is consistent with the anomalous scaling of the second-order structure function, which follows $S_2 \sim \log(r/\eta)$ for separations larger than the particle diameter, and indicates a loss of velocity correlations \citep{Olivieri_Cannon_Rosti_2022}. Further insight into the flow modulation is obtained from the scale-by-scale energy budget \citep{Pope_2000}. As the mass loading increases, the nonlinear energy transfer term weakens, while the fluid-solid interaction term becomes increasingly dominant. In the case of the heaviest particles, the energy balance is controlled primarily by the mechanical work of the fluid-solid interaction and by viscous dissipation across the whole spectrum, whereas the nonlinear interactions that characterise the classical Kolmogorov cascade are fully suppressed. The magnification of the fluid velocity skewness and kurtosis at the particle scales highlights an intensification of the small-scale activity. The PDFs of the invariants of the velocity gradient tensor indicate that increasing particle inertia enhances strain-dominated events at the expense of enstrophy, leading to a balance between axial and biaxial strain and between vortex stretching and compression. Notably, the tails of the PDFs of $R$ become increasingly pronounced, indicating stronger strain fields generated by the deformation of the streamlines around the solid spheres.

The breakdown of the large-scale structures observed in the visualizations of the dissipation and enstrophy fields, together with the rapid saturation of the structure functions at separations larger than the Kolmogorov scale, is reflected in both the individual and collective dynamics of the particles. The magnitude of the particle translational velocity decreases along with the fluid kinetic energy, whereas the fluid-particle velocity difference and the particle Reynolds number $\Rey_p$ increase with $M_p$. At the same time, the tails of the PDFs of $\Rey_p$ become less pronounced, indicating that an increasing number of particles undergoes a motion weakly correlated with the surrounding fluid as $M_p$ increases. Heavy particles are therefore subjected to a stronger nonlinear drag associated with the enhanced enstrophy and dissipation in their vicinity. As particle inertia increases, the suspension becomes progressively less sensitive to the carrier flow. The mean square displacement, which displays ballistic behavior at short times and approaches the Brownian limit at intermediate times, no longer reaches a clear asymptotic regime at large times.

We characterise the degree of clustering and the structure of the clusters using the Voronoi tessellation \citep{Monchaux_Bourgoin_Cartellier_2010}. The PDFs of the Voronoi volumes are broader for $\Psi_p=100$ and collapse on the PDF of a random Poisson process as the density ratio increases. The standard deviation of the volumes, which quantifies the degree of clustering, decreases for $\Psi_p>100$. Our results are complementary to those of \cite{Chiarini_Tandurella_Rosti_2025}, who reported an intensification of clustering when the density ratio increased from $\Psi_p=5$ to $100$ for a suspension of Kolmogorov-size particles of equal volume fraction. In this case, the Stokes number based on the Kolmogorov time scales has a narrow distribution around a mean $3.84$ and the average cluster length is $\approx20\eta$. No clusters can be identified in the heaviest-particle case, for which the PDF of the volumes collapses on that of the RPP. The PDFs of the cluster volumes show a power-law decay close to the exponents $-16/9$ and $-2$. The first is consistent with the fractal model of \cite{Yoshimoto_Goto_2007} and the second has been reported in numerical and experimental works \citep{Baker_Frankel_Mani_Coletti_2017,Hassaini_Petersen_Coletti_2023}. For low and moderate mass loadings, these power-law scalings indicate that cluster sizes span the inertial subrange. In the case $M_p=0.1$, the PDFs of the the invariant $Q$ sampled near the spheres are negatively skewed and the joint PDFs of $Q$ and the Voronoi volumes show that the small volumes correlate with negative $Q$, thus suggesting that the preferential sampling fosters cluster formation. Clusters shrink and become less and less populated as the density ratio increases and disappear completely when the heaviest particles ($M_p=0.6$) are considered.

Future work should address the role of multidisperse (i.e. non-uniform) suspensions of spherical and non-spherical Kolmogorov-size particles. In addition, further particle image velocimetry measurements with sub-Kolmogorov resolution \citep{Tanaka_Eaton_2010} are needed to resolve the multiscale structure of flows laden with Kolmogorov-size particles.


\begin{bmhead}[Acknowledgements]
The authors acknowledge the computational resources provided by the Scientific Computing \& Data Analysis Section of the Core Facilities at OIST and by HPCI under Research Project grants hp250021, hp250035, hp260009, and hp260019. The Authors would like to thank the anonymous Referees for their insightful comments. LF extends his thanks to Dr. Piyush Garg for his suggestions and to Mr. Simone Tandurella for valuable discussions regarding the Voronoi tessellation.
\end{bmhead}

\begin{bmhead}[Funding]
The research was supported by the Okinawa Institute of Science and Technology Graduate University (OIST) with subsidy funding to M.E.R. from the Cabinet Office, Government of Japan. M.E.R. also acknowledges funding from the Japan Society for the Promotion of Science (JSPS), grant 24K17210 and 24K00810.
\end{bmhead}

\begin{bmhead}[Declaration of interests]
The authors report no conflict of interest.
\end{bmhead}

\begin{bmhead}[Author ORCIDs] \\
  L. Foss\`{a}, https://orcid.org/0000-0001-7138-5903 \\ M. E. Rosti, https://orcid.org/0000-0002-9004-2292
\end{bmhead}

\bibliographystyle{jfm}
\bibliography{jfm}

@article{Ahmed_Khan_Bays-Muchmore_1993,
author = {Ahmed, A. and Khan, M. J. and Bays-Muchmore, B.},
title = {Experimental investigation of a three-dimensional bluff-body wake},
journal = {AIAA J.},
volume = {31},
number = {3},
pages = {559-563},
year = {1993},
doi = {10.2514/3.11365}
}

@article{Arnold_1965, 
  title={Sur la topologie des \'{e}coulements stationnaires des fluides parfaits}, 
  volume={261},  
  journal={C. R. Acad. Sci. Paris}, 
  author={Arnold, V. I.}, 
  year={1965}, 
  pages={17-20}
}

@article{Amsden_Harlow_1970,
title = {A simplified {MAC} technique for incompressible fluid flow calculations},
journal = {J. Comput. Phys.},
volume = {6},
number = {2},
pages = {322-325},
year = {1970},
issn = {0021-9991},
doi = {10.1016/0021-9991(70)90029-X},
author = {A. A. Amsden and F. H. Harlow}
}

@article{Ardekani_Brandt_2019, 
  title={Turbulence modulation in channel flow of finite-size spheroidal particles}, 
  volume={859}, 
  DOI={10.1017/jfm.2018.854}, 
  journal={J. Fluid Mech.}, 
  author={Ardekani, M. N. and Brandt, L.}, 
  year={2019}, 
  pages={887–901}
}

@article{Baker_Frankel_Mani_Coletti_2017, 
  title={Coherent clusters of inertial particles in homogeneous turbulence}, 
  volume={833}, 
  DOI={10.1017/jfm.2017.700}, 
  journal={J. Fluid Mech.}, 
  author={Baker, L. and Frankel, A. and Mani, A. and Coletti, F.}, 
  year={2017}, 
  pages={364–398}
}

@article{Balachandar_Eaton_2010,
  author = "Balachandar, S. and Eaton, J. K.",
  title = "Turbulent Dispersed Multiphase Flow", 
  journal= "Ann. Rev. Fluid Mech.",
  year = "2010",
  volume = "42",
  number = "Vol. 42, 2010",
  pages = "111-133",
  doi = "10.1146/annurev.fluid.010908.165243",
  publisher = "Annual Reviews",
  issn = "1545-4479",
  type = "Journal Article"
}

@article{Balachandar_Liu_Lakhote_2019,
  title = {Self-induced velocity correction for improved drag estimation in {Euler}-{Lagrange} point-particle simulations},
  journal = {J. Computat. Phys.},
  volume = {376},
  pages = {160-185},
  year = {2019},
  issn = {0021-9991},
  doi = {10.1016/j.jcp.2018.09.033},
  author = {S. Balachandar and K. Liu and M. Lakhote}
}

@article{Benzi_Ciliberto_Tripiccione_Baudet_Massaioli_Succi_1993,
  title = {Extended self-similarity in turbulent flows},
  author = {Benzi, R. and Ciliberto, S. and Tripiccione, R. and Baudet, C. and Massaioli, F. and Succi, S.},
  journal = {Phys. Rev. E},
  volume = {48},
  issue = {1},
  pages = {R29--R32},
  numpages = {0},
  year = {1993},
  month = {07},
  publisher = {American Physical Society},
  doi = {10.1103/PhysRevE.48.R29}
}

@article{Betchov_1956, 
  title={An inequality concerning the production of vorticity in isotropic turbulence}, 
  volume={1}, 
  DOI={10.1017/S0022112056000317}, 
  number={5}, 
  journal={J. Fluid Mech.}, 
  author={Betchov, R.}, 
  year={1956}, 
  pages={497–504}
}

@article{Biferale_Cencini_Lanotte_Vergni_2003,
    author = {Biferale, L. and Cencini, M. and Lanotte, A. S. and Vergni, D.},
    title = {Inverse velocity statistics in two-dimensional turbulence},
    journal = {Phys. Fluids},
    volume = {15},
    number = {4},
    pages = {1012-1020},
    year = {2003},
    month = {04},
    issn = {1070-6631},
    doi = {10.1063/1.1557527}
}

@article{Birnstiel_2024,
   author = "Birnstiel, T.",
   title = "Dust Growth and Evolution in Protoplanetary Disks", 
   journal= "Ann. Rev. Astron. Astrophys.",
   year = "2024",
   volume = "62",
   number = "Vol. 62, 2024",
   pages = "157-202",
   doi = "10.1146/annurev-astro-071221-052705",
   publisher = "Annual Reviews",
   issn = "1545-4282"
  }

@article{Borgnino_Arrieta_Boffetta_De-Lillo_Tuval_2019,
  author = {Borgnino, M. and Arrieta, J. and Boffetta, G. and De Lillo, F.  and Tuval, I.},
  title = {Turbulence induces clustering and segregation of non-motile, buoyancy-regulating phytoplankton},
  journal = {J. Royal Soc. Interface},
  volume = {16},
  number = {159},
  pages = {20190324},
  year = {2019},
  doi = {10.1098/rsif.2019.0324}
}

@article{Brandt_Coletti_2022,
   author = "Brandt, L. and Coletti, F.",
   title = "Particle-Laden Turbulence: Progress and Perspectives", 
   journal= "Ann. Rev. Fluid Mech.",
   year = "2022",
   volume = "54",
   number = "Vol. 54, 2022",
   pages = "159-189",
   doi = "10.1146/annurev-fluid-030121-021103",
   publisher = "Annual Reviews",
   issn = "1545-4479",
   type = "Journal Article"
  }

@article{Breugem_2012,
  title = {A second-order accurate immersed boundary method for fully resolved simulations of particle-laden flows},
  journal = {J. Comput. Phys.},
  volume = {231},
  number = {13},
  pages = {4469-4498},
  year = {2012},
  issn = {0021-9991},
  doi = {10.1016/j.jcp.2012.02.026},
  author = {W.-P. Breugem}
}

@article{Brown_Warhaft_Voth_2009,
  title = {Acceleration Statistics of Neutrally Buoyant Spherical Particles in Intense Turbulence},
  author = {Brown, R. D. and Warhaft, Z. and Voth, G. A.},
  journal = {Phys. Rev. Lett.},
  volume = {103},
  issue = {19},
  pages = {194501},
  numpages = {4},
  year = {2009},
  month = {Nov},
  publisher = {American Physical Society},
  doi = {10.1103/PhysRevLett.103.194501}
}

@article{Cannon_Olivieri_Rosti_2024,
  title = {Spheres and fibers in turbulent flows at various {Reynolds} numbers},
  author = {Cannon, I. and Olivieri, S. and Rosti, M. E.},
  journal = {Phys. Rev. Fluids},
  volume = {9},
  issue = {6},
  pages = {064301},
  numpages = {21},
  year = {2024},
  month = {06},
  publisher = {American Physical Society},
  doi = {10.1103/PhysRevFluids.9.064301}
}

@article{Cantwell_1992,
    author = {Cantwell, B. J.},
    title = {Exact solution of a restricted {Euler} equation for the velocity gradient tensor},
    journal = {Phys. Fluids A: Fluid Dyn.},
    volume = {4},
    number = {4},
    pages = {782-793},
    year = {1992},
    month = {04},
    issn = {0899-8213},
    doi = {10.1063/1.858295}
}

@article{Chiarini_Cannon_Rosti_2024,
  title = {Anisotropic Mean Flow Enhancement and Anomalous Transport of Finite-Size Spherical Particles in Turbulent Flows},
  author = {Chiarini, A. and Cannon, I. and Rosti, M. E.},
  journal = {Phys. Rev. Lett.},
  volume = {132},
  issue = {5},
  pages = {054005},
  numpages = {5},
  year = {2024},
  month = {02},
  publisher = {American Physical Society},
  doi = {10.1103/PhysRevLett.132.054005}
}

@article{Chiarini_Rosti_2024, 
  title={Finite-size inertial spherical particles in turbulence}, 
  volume={988}, 
  DOI={10.1017/jfm.2024.421}, 
  journal={J. Fluid Mech.}, 
  author={Chiarini, A. and Rosti, M. E.}, 
  year={2024},
  pages={A17}
}

@article{Chiarini_Tandurella_Rosti_2025, 
  title={Kolmogorov-size particles in homogeneous and isotropic turbulence}, 
  volume={1007}, 
  DOI={10.1017/jfm.2025.122}, 
  journal={J. Fluid Mech.}, 
  author={Chiarini, A. and Tandurella, S. and Rosti, M. E.}, 
  year={2025}, 
  pages={A81}
}

@Article{Chouippe_Uhlmann_2019,
    author={Chouippe, A. and Uhlmann, M.},
    title={On the influence of forced homogeneous-isotropic turbulence on the settling and clustering of finite-size particles},
    journal={Acta Mech.},
    year={2019},
    month={Feb},
    day={01},
    volume={230},
    number={2},
    pages={387-412},
    issn={1619-6937},
    doi={10.1007/s00707-018-2271-7}
}

@article{Codispoti_Meyer_Jenny_2025,
    author = {Codispoti, L. A. and Meyer, D. W. and Jenny, P.},
    title = {Dissecting inertial clustering and sling dynamics in high-{Reynolds} number particle-laden turbulence},
    journal = {Phys. Fluids},
    volume = {37},
    number = {1},
    pages = {015161},
    year = {2025},
    month = {01},
    issn = {1070-6631},
    doi = {10.1063/5.0244428}
}

@article{Costa_Boersma_Westerweel_Breugem_2015,
  title = {Collision model for fully resolved simulations of flows laden with finite-size particles},
  author = {Costa, P. and Boersma, B. J. and Westerweel, J. and Breugem, W.-P.},
  journal = {Phys. Rev. E},
  volume = {92},
  issue = {5},
  pages = {053012},
  numpages = {14},
  year = {2015},
  month = {Nov},
  publisher = {American Physical Society},
  doi = {10.1103/PhysRevE.92.053012}
}

@article{Costa_Brandt_Picano_2021, 
  title={Near-wall turbulence modulation by small inertial particles}, 
  volume={922}, 
  DOI={10.1017/jfm.2021.507}, 
  journal={J. Fluid Mech.}, 
  author={Costa, P. and Brandt, L. and Picano, F.}, 
  year={2021}, 
  pages={A9}
}

@book{Davidson_2015,
  title={Turbulence: An Introduction for Scientists and Engineers},
  author={Davidson, P.},
  isbn={9780198722595},
  lccn={2014954276},
  year={2015},
  publisher={Oxford University Press}
}

@Article{Elghobashi_1994,
author={Elghobashi, S.},
title={On predicting particle-laden turbulent flows},
journal={Appl. Sci. Res.},
year={1994},
month={Jun},
day={01},
volume={52},
number={4},
pages={309-329},
issn={1573-1987},
doi={10.1007/BF00936835}
}

@article{Fallon_Rogers_2002,
author={Fallon, T. and Rogers, C. B.},
title={Turbulence-induced preferential concentration of solid particles in microgravity conditions},
journal={Exp. Fluids},
year={2002},
month={08},
day={01},
volume={33},
number={2},
pages={233-241},
issn={1432-1114},
doi={10.1007/s00348-001-0394-3}
}

@article{Ferran_Machicoane_Aliseda_Obligado_2023, 
  title={An experimental study on the settling velocity of inertial particles in different homogeneous isotropic turbulent flows}, 
  volume={970}, 
  DOI={10.1017/jfm.2023.579}, 
  journal={J. Fluid Mech.}, 
  author={Ferran, A. and Machicoane, N. and Aliseda, A. and Obligado, M.}, 
  year={2023}, 
  pages={A23}
}

@article{Fiabane_Zimmermann_Volk_Pinton_Bourgoin_2012,
  title = {Clustering of finite-size particles in turbulence},
  author = {Fiabane, L. and Zimmermann, R. and Volk, R. and Pinton, J.-F. and Bourgoin, M.},
  journal = {Phys. Rev. E},
  volume = {86},
  issue = {3},
  pages = {035301},
  numpages = {4},
  year = {2012},
  month = {Sep},
  publisher = {American Physical Society},
  doi = {10.1103/PhysRevE.86.035301}
}

@article{Freund_2014,
   author = "Freund, J. B.",
   title = "Numerical Simulation of Flowing Blood Cells", 
   journal= "Ann. Rev. Fluid Mech.",
   year = "2014",
   volume = "46",
   number = "Volume 46, 2014",
   pages = "67-95",
   doi = "10.1146/annurev-fluid-010313-141349",
   publisher = "Annual Reviews",
   issn = "1545-4479"
  }

@article{Gao_Samtaney_Richter_2023, 
  title={Direct numerical simulation of particle-laden flow in an open channel at {$Re_{\tau}=5186$}}, 
  volume={957}, 
  DOI={10.1017/jfm.2023.26},  
  journal={J. Fluid Mech.}, 
  author={Gao, W. and Samtaney, R. and Richter, D. H.}, 
  year={2023}, 
  pages={A3}
}

@article{Good_Ireland_Bewley_Bodenschatz_Collins_Warhaft_2014, 
  title={Settling regimes of inertial particles in isotropic turbulence}, 
  volume={759}, 
  DOI={10.1017/jfm.2014.602}, 
  journal={J. Fluid Mech.}, 
  author={Good, G. H. and Ireland, P. J. and Bewley, G. P. and Bodenschatz, E. and Collins, L. R. and Warhaft, Z.}, 
  year={2014}, 
  pages={R3}
}

@article{Gore_Crowe_1989,
  title = {Effect of particle size on modulating turbulent intensity},
  journal = {Int. J. Multiph. Flow},
  volume = {15},
  number = {2},
  pages = {279-285},
  year = {1989},
  issn = {0301-9322},
  doi = {10.1016/0301-9322(89)90076-1},
  author = {R.A. Gore and C.T. Crowe}
}

@article{Goto_Vassilicos_2006,
    author = {Goto, S. and Vassilicos, J. C.},
    title = {Self-similar clustering of inertial particles and zero-acceleration points in fully developed two-dimensional turbulence},
    journal = {Phys. Fluids},
    volume = {18},
    number = {11},
    pages = {115103},
    year = {2006},
    month = {11},
    issn = {1070-6631},
    doi = {10.1063/1.2364263}
}

@article{Goto_Vassilicos_2008,
  title = {Sweep-Stick Mechanism of Heavy Particle Clustering in Fluid Turbulence},
  author = {Goto, S. and Vassilicos, J. C.},
  journal = {Phys. Rev. Lett.},
  volume = {100},
  issue = {5},
  pages = {054503},
  numpages = {4},
  year = {2008},
  month = {02},
  publisher = {American Physical Society},
  doi = {10.1103/PhysRevLett.100.054503}
}

@article{Gualtieri_Battista_Salvadore_Casciola_2023, 
  title={Effect of {Stokes} number and particle-to-fluid density ratio on turbulence modification in channel flows}, 
  volume={974}, 
  DOI={10.1017/jfm.2023.851}, 
  journal={J. Fluid Mech.}, 
  author={Gualtieri, P. and Battista, F. and Salvadore, F. and Casciola, C.M.}, 
  year={2023}, 
  pages={A26}
}

@article{Hassaini_Coletti_2022, 
  title={Scale-to-scale turbulence modification by small settling particles}, 
  volume={949}, 
  DOI={10.1017/jfm.2022.762}, 
  journal={J. Fluid Mech.}, 
  author={Hassaini, R. and Coletti, F.}, 
  year={2022}, 
  pages={A30}
}

@article{Hassaini_Petersen_Coletti_2023, 
  title={Effect of two-way coupling on clustering and settling of heavy particles in homogeneous turbulence}, 
  volume={976}, 
  DOI={10.1017/jfm.2023.896}, 
  journal={J. Fluid Mech.}, 
  author={Hassaini, R. and Petersen, A. J. and Coletti, F.}, 
  year={2023}, 
  pages={A12}
}

@article{Hoque_Mitra_Sathe_Joshi_Evans_2016,
title = {Experimental investigation on modulation of homogeneous and isotropic turbulence in the presence of single particle using time-resolved {PIV}},
journal = {Chem. Eng. Sci.},
volume = {153},
pages = {308-329},
year = {2016},
issn = {0009-2509},
doi = {10.1016/j.ces.2016.07.026},
author = {M. M. Hoque and S. Mitra and M. J. Sathe and J. B. Joshi and G. M. Evans}
}

@article{Hori_Takagi_Rosti_2022,
title = {An {Eulerian}-based immersed boundary method for particle suspensions with implicit lubrication model},
journal = {Comput. Fluids},
volume = {236},
pages = {105278},
year = {2022},
issn = {0045-7930},
doi = {10.1016/j.compfluid.2021.105278},
author = {N. Hori and M. E. Rosti and S. Takagi}
}

@article{Hwang_Eaton_2006_gravity, 
title={Homogeneous and isotropic turbulence modulation by small heavy (${St}\sim 50$) particles}, 
volume={564}, 
DOI={10.1017/S0022112006001431}, 
journal={J. Fluid Mech.}, 
author={Hwang, W. and Eaton, J. K.}, 
year={2006}, 
pages={361–393}
}

@article{Hwang_Eaton_2006_nogravity,
title = {Turbulence attenuation by small particles in the absence of gravity},
journal = {Int. J. Multiph. Flow},
volume = {32},
number = {12},
pages = {1386-1396},
year = {2006},
issn = {0301-9322},
doi = {10.1016/j.ijmultiphaseflow.2006.06.008},
author = {W. Hwang and J.K. Eaton}
}

@article{Jenny_Roekaerts_Beishuizen_2012,
title = {Modeling of turbulent dilute spray combustion},
journal = {Prog. Energy Combust. Sci.},
volume = {38},
number = {6},
pages = {846-887},
year = {2012},
issn = {0360-1285},
doi = {10.1016/j.pecs.2012.07.001},
author = {Jenny, P. and Roekaerts, D. and Beishuizen, N.}
}

@article{Jiang_Brandt_Xu_Zhao_2025, 
  title={Pseudo-turbulence induced by settling spheroids in a quiescent fluid}, 
  volume={1011}, 
  DOI={10.1017/jfm.2025.398}, 
  journal={J. Fluid Mech.}, 
  author={Jiang, X. and Brandt, L. and Xu, C. and Zhao, L.}, 
  year={2025}, 
  pages={A22}
}

@article{Jiang_Mirzareza_Crialesi-Esposito_Brandt_2026, 
title={Turbulence modulation by {Taylor}-scale settling particles}, 
volume={1034}, 
DOI={10.1017/jfm.2026.11476}, 
journal={J. Fluid Mech.}, 
author={Jiang, X. and Mirzareza, S. and Crialesi Esposito, M. and Brandt, L.}, 
year={2026}, 
pages={A27}
}

@article{Kajishima_Takiguchi_Hamasaki_Miyake_2001,
  title={Turbulence Structure of Particle-Laden Flow in a Vertical Plane Channel Due to Vortex Shedding},
  author={Kajishima, T. and Takiguchi, S. and Hamasaki, H. and Miyake, Y.},
  journal={JSME Int. J. Ser. B Fluids Therm. Eng.},
  volume={44},
  number={4},
  pages={526-535},
  year={2001},
  doi={10.1299/jsmeb.44.526}
}

@article{Kempe_Frohlich_2012,
title = {An improved immersed boundary method with direct forcing for the simulation of particle laden flows},
journal = {J. Comput. Phys.},
volume = {231},
number = {9},
pages = {3663-3684},
year = {2012},
issn = {0021-9991},
doi = {10.1016/j.jcp.2012.01.021},
author = {Kempe, T. and Fröhlich, J.}
}

@article{Kim_Moin_Moser_1987, 
  title={Turbulence statistics in fully developed channel flow at low {Reynolds} number}, 
  volume={177}, 
  DOI={10.1017/S0022112087000892}, 
  journal={J. Fluid Mech.}, 
  author={Kim, J. and Moin, P. and Moser, R.}, 
  year={1987}, 
  pages={133–166}
}

@article{Lee_Raj-Mohan_Byeon_Lim_Hong_2013,
  author = {Lee, B.-K. and B. Raj Mohan, B. and Byeon, S.-H. and Lim, K.-S. and Hong, E.-P.},
  title = {Evaluating the performance of a turbulent wet scrubber for scrubbing particulate matter},
  journal = {J. Air Waste Manag. Assoc.},
  volume = {63},
  number = {5},
  pages = {499--506},
  year = {2013},
  publisher = {Taylor \& Francis},
  doi = {10.1080/10962247.2012.738626}
}

@article{Lucci_Ferrante_Elghobashi_2010, 
  title={Modulation of isotropic turbulence by particles of {Taylor} length-scale size}, 
  volume={650}, 
  DOI={10.1017/S0022112009994022}, 
  journal={J. Fluid Mech.}, 
  author={Lucci, F. and Ferrante, A. and Elghobashi, S.}, 
  year={2010}, 
  pages={5–55}
}

@article{Lu_Lazar_Rycroft_2023,
  title = {An extension to {Voro++} for multithreaded computation of {Voronoi} cells},
  journal = {Comput. Phys. Commun.},
  volume = {291},
  pages = {108832},
  year = {2023},
  issn = {0010-4655},
  doi = {10.1016/j.cpc.2023.108832},
  author = {J. Lu and E. A. Lazar and C. H. Rycroft}
}

@article{Luo_Wang_Li_Tan_Fan_2017,
    author = {Luo, K. and Wang, Z. and Li, D. and Tan, J. and Fan, J.},
    title = {Fully resolved simulations of turbulence modulation by high-inertia particles in an isotropic turbulent flow},
    journal = {Phys. Fluids},
    volume = {29},
    number = {11},
    pages = {113301},
    year = {2017},
    month = {11},
    issn = {1070-6631},
    doi = {10.1063/1.4997731}
}

@article{Marchioli_Bourgoin_Coletti_Fox_Magnaudet_Reeks_Simonin_Sommerfeld_Toschi_Wang_Balachandar_2025,
title = {Particle-laden flows},
journal = {Int. J. Multiph. Flow},
volume = {191},
pages = {105291},
year = {2025},
issn = {0301-9322},
doi = {10.1016/j.ijmultiphaseflow.2025.105291},
author = {C. Marchioli and M. Bourgoin and F. Coletti and R. O. Fox and J. Magnaudet and M. Reeks and O. Simonin and M. Sommerfeld and F. Toschi and L.-P. Wang and S. Balachandar},
}

@article{Marchioli_Rosti_Verhille_2025,
  author = {Marchioli, C. and Rosti, M. E. and Verhille, G.},
  title = {Flexible Fibers in Turbulence}, 
  journal= {Ann. Rev. Fluid Mech.},
  year = {2026},
  volume = {58},
  number = {Vol. 58, 2026},
  pages = {167-192},
  doi = {10.1146/annurev-fluid-112723-050451},
  publisher = {Annual Reviews}
}

@article{Martinez-Mercado_Prakash_Tagawa_Sun_Lohse_2012,
    author = {Mart\'inez Mercado, J. and Prakash, V. N. and Tagawa, Y. and Sun, C. and Lohse, D.},
    title = {Lagrangian statistics of light particles in turbulence},
    journal = {Phys. Fluids},
    volume = {24},
    number = {5},
    pages = {055106},
    year = {2012},
    month = {05},
    issn = {1070-6631},
    doi = {10.1063/1.4719148}
}

@article{Matsuda_Yoshimatsu_Schneider_2024,
  title = {Heavy Particle Clustering in Inertial Subrange of High--{Reynolds} Number Turbulence},
  author = {Matsuda, K. and Yoshimatsu, K. and Schneider, K.},
  journal = {Phys. Rev. Lett.},
  volume = {132},
  issue = {23},
  pages = {234001},
  numpages = {6},
  year = {2024},
  month = {06},
  publisher = {American Physical Society},
  doi = {10.1103/PhysRevLett.132.234001}
}

@article{Maxey_Riley_1983,
    author = {Maxey, M. R. and Riley, J. J.},
    title = {Equation of motion for a small rigid sphere in a nonuniform flow},
    journal = {Phys. Fluids},
    volume = {26},
    number = {4},
    pages = {883-889},
    year = {1983},
    month = {04},
    issn = {0031-9171},
    doi = {10.1063/1.864230}
}

@article{Maxey_1987, 
  title={The gravitational settling of aerosol particles in homogeneous turbulence and random flow fields}, 
  volume={174}, 
  DOI={10.1017/S0022112087000193}, 
  journal={J. Fluid Mech.}, 
  author={Maxey, M. R.}, 
  year={1987}, 
  pages={441–465}
}

@article{Mellado_2017,
   author = "Mellado, J. P.",
   title = "Cloud-Top Entrainment in Stratocumulus Clouds", 
   journal= "Ann. Rev. Fluid Mech.",
   year = "2017",
   volume = "49",
   number = "Volume 49, 2017",
   pages = "145-169",
   doi = "10.1146/annurev-fluid-010816-060231",
   publisher = "Annual Reviews",
   issn = "1545-4479"
}

@article{Meneveau_2011,
  author = "Meneveau, C.",
  title = "Lagrangian Dynamics and Models of the Velocity Gradient Tensor in Turbulent Flows", 
  journal= "Ann. Rev. Fluid Mech.",
  year = "2011",
  volume = "43",
  number = "Vol. 43, 2011",
  pages = "219-245",
  doi = "10.1146/annurev-fluid-122109-160708",
  publisher = "Annual Reviews",
  issn = "1545-4479"
}

@article{Monchaux_Bourgoin_Cartellier_2010,
  author = {Monchaux, R. and Bourgoin, M. and Cartellier, A.},
  title = {Preferential concentration of heavy particles: A {Voronoi} analysis},
  journal = {Phys. Fluids},
  volume = {22},
  number = {10},
  pages = {103304},
  year = {2010},
  month = {10},
  issn = {1070-6631},
  doi = {10.1063/1.3489987}
}

@article{Monchaux_2012,
doi = {10.1088/1367-2630/14/9/095013},
year = {2012},
month = {sep},
publisher = {IOP Publishing},
volume = {14},
number = {9},
pages = {095013},
author = {Monchaux, R.},
title = {Measuring concentration with {Voronoi} diagrams: the study of possible biases},
journal = {New J. Phys.}}

@article{Monti_Rathee_Shen_Rosti_2021,
    author = {Monti, A. and Rathee, V. and Shen, A. Q. and Rosti, M. E.},
    title = {A fast and efficient tool to study the rheology of dense suspensions},
    journal = {Phys. Fluids},
    volume = {33},
    number = {10},
    pages = {103314},
    year = {2021},
    month = {10},
    issn = {1070-6631},
    doi = {10.1063/5.0065655}
}

@article{Mortimer_Fairweather_2020,
    author = {Mortimer, L. F. and Fairweather, M.},
    title = {Density ratio effects on the topology of coherent turbulent structures in two-way coupled particle-laden channel flows},
    journal = {Phys. Fluids},
    volume = {32},
    number = {10},
    pages = {103302},
    year = {2020},
    month = {10},
    issn = {1070-6631},
    doi = {10.1063/5.0017458},
    url = {10.1063/5.0017458}
}

@article{Obligado_Teitelbaum_Cartellier_Mininni_Bourgoin_2014,
  author = {M. Obligado and T. Teitelbaum and A. Cartellier and P. Mininni and M. Bourgoin},
  title = {Preferential concentration of heavy particles in turbulence},
  journal = {J. Turbul.},
  volume = {15},
  number = {5},
  pages = {293--310},
  year = {2014},
  publisher = {Taylor \& Francis},
  doi = {10.1080/14685248.2014.897710}
}

@article{Olivieri_Cannon_Rosti_2022, 
  title={The effect of particle anisotropy on the modulation of turbulent flows}, 
  volume={950}, 
  DOI={10.1017/jfm.2022.832}, 
  journal={J. Fluid Mech.}, 
  author={Olivieri, S. and Cannon, I. and Rosti, M. E.}, 
  year={2022}, 
  pages={R2}
}

@article{Oka_Goto_2022, 
  title={Attenuation of turbulence in a periodic cube by finite-size spherical solid particles}, 
  volume={949}, 
  DOI={10.1017/jfm.2022.787}, 
  journal={J. Fluid Mech.}, 
  author={Oka, S. and Goto, S.}, 
  year={2022}, 
  pages={A45}
}

@article{Poelma_Westerweel_Ooms_2007, 
  title={Particle–fluid interactions in grid-generated turbulence}, 
  volume={589}, 
  DOI={10.1017/S0022112007007793}, 
  journal={J. Fluid Mech.}, 
  author={Poelma, C. and Westerweel, J. and Ooms, G.}, 
  year={2007}, 
  pages={315–351}
}

@book{Pope_2000,
  title={Turbulent Flows},
  author={Pope, S.B.},
  isbn={9780521598866},
  lccn={99044583},
  year={2000},
  publisher={Cambridge University Press}
}

@article{Prakash_Martinez-Mercado_van-Wijngaarden_Mancilla_Tagawa_Lohse_Sun_2016, 
title={Energy spectra in turbulent bubbly flows}, 
volume={791}, 
DOI={10.1017/jfm.2016.49}, 
journal={J. Fluid Mech.}, 
author={Prakash, V. N. and Mart\'{i}nez Mercado, J. and van Wijngaarden, L. and Mancilla, E. and Tagawa, Y. and Lohse, D. and Sun, C.}, 
year={2016}, 
pages={174–190}
}

@article{Ramirez_Burlot_Zamansky_Bois_Risso_2024,
title = {Spectral analysis of dispersed multiphase flows in the presence of fluid interfaces},
journal = {Int. J. Multiph. Flow},
volume = {177},
pages = {104860},
year = {2024},
issn = {0301-9322},
doi = {10.1016/j.ijmultiphaseflow.2024.104860},
author = {G. Ramirez and A. Burlot and R. Zamansky and G. Bois and F. Risso}
}

@article{Rose_Durant_2009,
title = {Fine ash content of explosive eruptions},
journal = {J. Volcanol. Geotherm. Res.},
volume = {186},
number = {1},
pages = {32-39},
year = {2009},
issn = {0377-0273},
doi = {10.1016/j.jvolgeores.2009.01.010},
author = {W. I. Rose and A. J. Durant}
}

@article{Rostami_Li_Kheirkhah_2024, 
  title={Three-dimensional clustering characteristics of large-{Stokes}-number dilute sprays interacting with turbulent swirling co-flows}, 
  volume={999}, 
  DOI={10.1017/jfm.2024.926}, 
  journal={J. Fluid Mech.}, 
  author={Rostami, A. and Li, R. and Kheirkhah, S.}, 
  year={2024}, 
  pages={A73}
}

@article{Rosti_2026,
doi = {10.1088/1873-7005/ae4fee},
year = {2026},
month = {03},
publisher = {IOP Publishing},
volume = {58},
number = {2},
pages = {021401},
author = {Rosti, M. E.},
title = {Simulating laminar and turbulent multiphase flows with {Fujin}},
journal = {Fluid Dyn. Res.}
}

@article{Rosti_Brandt_2020,
  title = {Increase of turbulent drag by polymers in particle suspensions},
  author = {Rosti, M. E. and Brandt, L.},
  journal = {Phys. Rev. Fluids},
  volume = {5},
  issue = {4},
  pages = {041301},
  numpages = {9},
  year = {2020},
  month = {Apr},
  publisher = {American Physical Society},
  doi = {10.1103/PhysRevFluids.5.041301}
}

@article{Salazar_De-Jong_Cao_Woodward_Meng_Collins_2008, 
  title={Experimental and numerical investigation of inertial particle clustering in isotropic turbulence}, 
  volume={600}, 
  DOI={10.1017/S0022112008000372}, 
  journal={J. Fluid Mech.}, 
  author={Salazar, J. P. L. C. and De Jong, J. and Cao, L. and Woodward, S. H. and Meng, H. and Collins, L. R.}, 
  year={2008}, 
  pages={245–256}
}

@article{Saw_Shaw_Ayyalasomayajula_Chuang_Gylfason_2008,
  title = {Inertial Clustering of Particles in High-{Reynolds}-Number Turbulence},
  author = {Saw, E. W. and Shaw, R. A. and Ayyalasomayajula, S. and Chuang, P. Y. and Gylfason, \'A.},
  journal = {Phys. Rev. Lett.},
  volume = {100},
  issue = {21},
  pages = {214501},
  numpages = {4},
  year = {2008},
  month = {05},
  publisher = {American Physical Society},
  doi = {10.1103/PhysRevLett.100.214501}
}

@article{Schiller_Naumann_1933, 
  title={Uber die grundlegenden berechnungen bei der schwerkraftaufbereitung}, 
  volume={77}, 
  journal={Ztg. Ver. 	Dtsch. Ing.}, 
  author={Schiller, L. and Naumann, A.}, 
  year={1933}, 
  pages={318–320}
}

@article{Schneiders_Meinke_Schröder_2017, 
  title={Direct particle–fluid simulation of {Kolmogorov}-length-scale size particles in decaying isotropic turbulence}, 
  volume={819}, 
  DOI={10.1017/jfm.2017.171}, 
  journal={J. Fluid Mech.}, 
  author={Schneiders, L. and Meinke, M. and Schr\"{o}der, W.}, 
  year={2017}, 
  pages={188–227}
}

@article{Shen_Peng_Lu_Wang_2024,
  title = {The influence of particle density and diameter on the interactions between the finite-size particles and the turbulent channel flow},
  journal = {Int. J. Multiph. Flow},
  volume = {170},
  pages = {104659},
  year = {2024},
  issn = {0301-9322},
  doi = {10.1016/j.ijmultiphaseflow.2023.104659},
  author = {Shen, J. and Peng, C. and Lu, Z. and Wang, L.-P.},
}

@article{Squires_Eaton_1990,
  author = {Squires, K. D. and Eaton, J. K.},
  title = {Particle response and turbulence modification in isotropic turbulence},
  journal = {Phys. Fluids A: Fluid Dyn.},
  volume = {2},
  number = {7},
  pages = {1191-1203},
  year = {1990},
  month = {07},
  issn = {0899-8213},
  doi = {10.1063/1.857620}
}

@article{Sugathapala_Capuano_Brandt_Iudicone_Sardina_2025,
title = {Vertical transport of buoyant microplastic particles in the ocean: The role of turbulence and biofouling},
journal = {Environ. Pollut.},
volume = {369},
pages = {125819},
year = {2025},
issn = {0269-7491},
doi = {10.1016/j.envpol.2025.125819},
author = {T. M. Sugathapala and T. Capuano and L. Brandt and D. Iudicone and G. Sardina}
}

@article{Sumbekova_Cartellier_Aliseda_Bourgoin_2017,
  title = {Preferential concentration of inertial sub-{Kolmogorov} particles: The roles of mass loading of particles, {Stokes} numbers, and {Reynolds} numbers},
  author = {Sumbekova, S. and Cartellier, A. and Aliseda, A. and Bourgoin, M.},
  journal = {Phys. Rev. Fluids},
  volume = {2},
  issue = {2},
  pages = {024302},
  numpages = {19},
  year = {2017},
  month = {Feb},
  publisher = {American Physical Society},
  doi = {10.1103/PhysRevFluids.2.024302}
}

@article{Tagawa_Martinez-Mercado_Prakash_Calzavarini_Sun_Lohse_2012, 
  title={Three-dimensional {Lagrangian} {Voronoi} analysis for clustering of particles and bubbles in turbulence}, 
  volume={693}, 
  DOI={10.1017/jfm.2011.510}, 
  journal={J. Fluid Mech.}, 
  author={Tagawa, Y. and Mart\`inez Mercado, J. and Prakash, Vivek N. and Calzavarini, E. and Sun, C. and Lohse, D.}, 
  year={2012}, 
  pages={201–215}
}

@article{Tanaka_Eaton_2008,
  title = {Classification of Turbulence Modification by Dispersed Spheres Using a Novel Dimensionless Number},
  author = {Tanaka, T. and Eaton, J. K.},
  journal = {Phys. Rev. Lett.},
  volume = {101},
  issue = {11},
  pages = {114502},
  numpages = {4},
  year = {2008},
  month = {09},
  publisher = {American Physical Society},
  doi = {10.1103/PhysRevLett.101.114502}
}

@article{Tanaka_Eaton_2010, 
  title={Sub-{Kolmogorov} resolution partical image velocimetry measurements of particle-laden forced turbulence}, 
  volume={643}, 
  DOI={10.1017/S0022112009992023}, 
  journal={J. Fluid Mech.}, 
  author={Tanaka, T. and Eaton, J. K.}, 
  year={2010}, 
  pages={177–206}
}

@Article{Tavanashad_Passalacqua_Fox_Subramaniam_2019,
  author={Tavanashad, V.
  and Passalacqua, A.
  and Fox, R. O.
  and Subramaniam, S.},
  title={Effect of density ratio on velocity fluctuations in dispersed multiphase flow from simulations of finite-size particles},
  journal={Acta Mech.},
  year={2019},
  month={02},
  day={01},
  volume={230},
  number={2},
  pages={469-484},
  issn={1619-6937},
  doi={10.1007/s00707-018-2267-3}
}

@article{Tome_Duffy_McKee_1996,
title = {A numerical technique for solving unsteady non-{Newtonian} free surface flows},
journal = {J. Non-Newton. Fluid Mech.},
volume = {62},
number = {1},
pages = {9-34},
year = {1996},
issn = {0377-0257},
doi = {10.1016/0377-0257(95)01391-1},
author = {M. F. Tom\'{e} and B. Duffy and S. McKee}
}

@article{Tsuji_Morikawa_1982, 
title={{LDV} measurements of an air-solid two-phase flow in a horizontal pipe}, 
volume={120}, 
DOI={10.1017/S002211208200281X}, 
journal={J. Fluid Mech.}, 
author={Tsuji, Y. and Morikawa, Y.}, 
year={1982}, 
pages={385–409}
}

@article{Tsuji_Kawaguchi_Tanaka_1993,
title = {Discrete particle simulation of two-dimensional fluidized bed},
journal = {Powder Tech.},
volume = {77},
number = {1},
pages = {79-87},
year = {1993},
issn = {0032-5910},
doi = {10.1016/0032-5910(93)85010-7},
author = {Y. Tsuji and T. Kawaguchi and T. Tanaka}
}

@article{Uhlmann_2005,
  title = {An immersed boundary method with direct forcing for the simulation of particulate flows},
  journal = {J. Comput. Phys.},
  volume = {209},
  number = {2},
  pages = {448-476},
  year = {2005},
  issn = {0021-9991},
  doi = {10.1016/j.jcp.2005.03.017},
  author = {Uhlmann, M.}
}

@article{Uhlmann_2008,
  author = {Uhlmann, M.},
  title = {Interface-resolved direct numerical simulation of vertical particulate channel flow in the turbulent regime},
  journal = {Phys. Fluids},
  volume = {20},
  number = {5},
  pages = {053305},
  year = {2008},
  month = {05},
  issn = {1070-6631},
  doi = {10.1063/1.2912459}
}

@article{Uhlmann_Chouippe_2017, 
title={Clustering and preferential concentration of finite-size particles in forced homogeneous-isotropic turbulence}, 
volume={812}, 
DOI={10.1017/jfm.2016.826}, 
journal={J. Fluid Mech.}, 
author={Uhlmann, M. and Chouippe, A.}, 
year={2017}, 
pages={991–1023}
}

@article{Volk_Calzavarini_Verhille_Lohse_Mordant_Pinton_Toschi_2008,
title = {Acceleration of heavy and light particles in turbulence: Comparison between experiments and direct numerical simulations},
journal = {Phys. D: Nonlinear Phenom.},
volume = {237},
number = {14},
pages = {2084-2089},
year = {2008},
issn = {0167-2789},
doi = {10.1016/j.physd.2008.01.016},
author = {R. Volk and E. Calzavarini and G. Verhille and D. Lohse and N. Mordant and J.-F. Pinton and F. Toschi}
}

@article{Voth_La-Porta_Crawford_Alexander_Bodenschatz_2002, 
  title={Measurement of particle accelerations in fully developed turbulence}, 
  volume={469}, 
  DOI={10.1017/S0022112002001842}, 
  journal={J. Fluid Mech.}, 
  author={Voth, G. A. and La Porta, A. and Crawford, A. M. and Alexander, J. and Bodenschatz, E.}, 
  year={2002}, 
  pages={121–160}
}

@article{van-Wachem_Elmestikawy_Chandran_Hausmann_2025, 
title={A new paradigm for computing hydrodynamic forces on particles in {Euler-Lagrange} point-particle simulations}, 
volume={1018}, 
DOI={10.1017/jfm.2025.10526}, 
journal={J. Fluid Mech.}, 
author={van Wachem, B. and Elmestikawy, H. and Chandran, A. and Hausmann, M.}, 
year={2025}, 
pages={A41}}

@article{Wang_Ayala_Gao_Andersen_Mathews_2014,
title = {Study of forced turbulence and its modulation by finite-size solid particles using the lattice {Boltzmann} approach},
journal = {Comput. \& Math. Appl.},
volume = {67},
number = {2},
pages = {363-380},
year = {2014},
issn = {0898-1221},
doi = {10.1016/j.camwa.2013.04.001},
author = {L.-P. Wang and O. Ayala and H. Gao and C. Andersen and K. L. Mathews}
}

@article{Yang_Shy_2005, 
  title={Two-way interaction between solid particles and homogeneous air turbulence: particle settling rate and turbulence modification measurements}, 
  volume={526}, 
  DOI={10.1017/S0022112004002861}, 
  journal={J. Fluid Mech.}, 
  author={Yang, T. S. and Shy, S. S.},      
  year={2005}, 
  pages={171–216}
}

@article{Yoshimoto_Goto_2007, 
  title={Self-similar clustering of inertial particles in homogeneous turbulence}, 
  volume={577}, 
  DOI={10.1017/S0022112007004946}, 
  journal={J. Fluid Mech.}, 
  author={Yoshimoto, H. and Goto, S.}, 
  year={2007}, 
  pages={275–286}
}

@article{Yu_Lin_Shao_Wang_2017,
  title = {Effects of particle-fluid density ratio on the interactions between the turbulent channel flow and finite-size particles},
  author = {Yu, Z. and Lin, Z. and Shao, X. and Wang, L.-P.},
  journal = {Phys. Rev. E},
  volume = {96},
  issue = {3},
  pages = {033102},
  numpages = {15},
  year = {2017},
  month = {Sep},
  publisher = {American Physical Society},
  doi = {10.1103/PhysRevE.96.033102}
}

\end{document}